\documentclass[journal]{IEEEtran}
\usepackage{amsmath,amsfonts}
\usepackage{algorithmic}
\usepackage{algorithm}
\usepackage{array}
\usepackage[caption=false,font=normalsize,labelfont=sf,textfont=sf]{subfig}
\usepackage{textcomp}
\usepackage{stfloats}
\usepackage{url}
\usepackage{verbatim}
\usepackage{graphicx}
\usepackage{cite}
\usepackage[hidelinks]{hyperref}
\usepackage{orcidlink}
\usepackage{acronym}
\usepackage{bibentry}
\usepackage{etoolbox}
\usepackage{multirow}
\usepackage{arydshln}
\usepackage{adjustbox}

\usetikzlibrary{backgrounds}

\makeatletter
\newcommand{\citetal}[2]{#1~\emph{et al.}~\cite{#2}}
\makeatother

\acrodef{ac}[AC]{Affective Computing}
\acrodef{acc}[ACC]{Accelerometer}
\acrodef{adr}[ADR]{Accelerometry-Derived Respiration}
\acrodef{ai}[AI]{Artificial Intelligence}
\acrodef{ans}[ANS]{Autonomic Nervous System}
\acrodef{ao}[AO]{Aortic valve Opening}
\acrodef{bb}[BB]{Breath to Breath}
\acrodef{bh3}[BH3]{BioHarness 3}
\acrodef{br}[BR]{Breath Rate}
\acrodef{bvp}[BVP]{Blood Volume Pulse}
\acrodef{cnn}[CNN]{Convolutional Neural Network}
\acrodef{cv}[CV]{Cross-Validation}
\acrodef{dt}[DT]{Decision Tree}
\acrodef{ecgy}[ECG]{Electrocardiography}
\acrodef{eda}[EDA]{Electrodermal Activity}
\acrodef{eegy}[EEG]{Electroencephalography}
\acrodef{emgy}[EMG]{Electromyography}
\acrodef{eogy}[EOG]{Electrooculography}
\acrodef{er}[ER]{Emotion Recognition}
\acrodef{gui}[GUI]{Graphical User Interface}
\acrodef{gyr}[GYRO]{Gyroscope}
\acrodef{hr}[HR]{Heart Rate}
\acrodef{hrv}[HRV]{Heart Rate Variability}
\acrodef{ibi}[IBI]{Inter-beat Interval}
\acrodef{imu}[IMU]{Inertial Measurement Unit}
\acrodef{knn}[k-NN]{k-Nearest Neighbors}
\acrodef{lr}[LR]{Logistic Regression}
\acrodef{lstm}[LSTM]{Long Short-Term Memory}
\acrodef{ml}[ML]{Machine Learning}
\acrodef{mlp}[MLP]{Multi-Layer Perceptron}
\acrodef{mqtt}[MQTT]{Message Queuing Telemetry Transport}
\acrodef{nb}[NB]{Naïve Bayes}
\acrodef{rf}[RF]{Radio Frequency}
\acrodef{rip}[RIP]{Respiratory Inductance Plethysmography}
\acrodef{rmanova}[rmANOVA]{repeated-measures ANalysis Of VAriance}
\acrodef{rr}[RR]{R wave to R wave}
\acrodef{rsp}[RSP]{respiration}
\acrodef{rtos}[RTOS]{Real-Time Operating System}
\acrodef{sam}[SAM]{Self-Assessment Manikins}
\acrodef{scg}[SCG]{seismocardiography}
\acrodef{skt}[SKT]{Skin Temperature}
\acrodef{snr}[SNR]{Signal-to-Noise Ratio}
\acrodef{stb}[STb]{SensorTile.box}
\acrodef{svm}[SVM]{Support Vector Machine}
\acrodef{vad}[VAD]{Voice Activity Detection}

\newacroindefinite{ac}{an}{an}
\newacroindefinite{acc}{an}{an}
\newacroindefinite{adr}{an}{an}
\newacroindefinite{ai}{an}{an}
\newacroindefinite{ans}{an}{an}
\newacroindefinite{ao}{an}{an}
\newacroindefinite{bb}{a}{a}
\newacroindefinite{bh3}{a}{a}
\newacroindefinite{br}{a}{a}
\newacroindefinite{bvp}{a}{a}
\newacroindefinite{cnn}{a}{a}
\newacroindefinite{ecgy}{an}{an}
\newacroindefinite{eda}{an}{an}
\newacroindefinite{eegy}{an}{an}
\newacroindefinite{emgy}{an}{an}
\newacroindefinite{eogy}{an}{an}
\newacroindefinite{er}{an}{an}
\newacroindefinite{gui}{a}{a}
\newacroindefinite{gyr}{a}{a}
\newacroindefinite{hr}{an}{a}
\newacroindefinite{hrv}{an}{a}
\newacroindefinite{ibi}{an}{an}
\newacroindefinite{imu}{an}{an}
\newacroindefinite{lr}{an}{a}
\newacroindefinite{ml}{an}{a}
\newacroindefinite{mqtt}{an}{a}
\newacroindefinite{nb}{an}{a}
\newacroindefinite{rf}{an}{a}
\newacroindefinite{rip}{an}{a}
\newacroindefinite{rmanova}{an}{a}
\newacroindefinite{rr}{an}{an}
\newacroindefinite{rsp}{an}{a}
\newacroindefinite{rtos}{an}{a}
\newacroindefinite{scg}{an}{a}
\newacroindefinite{skt}{an}{a}
\newacroindefinite{snr}{an}{a}
\newacroindefinite{stb}{an}{a}
\newacroindefinite{svm}{an}{a}
\newacroindefinite{vad}{a}{a}

\newcolumntype{L}[1]{>{\raggedright\arraybackslash\hspace{0pt}}p{#1}}
\newcolumntype{C}[1]{>{\centering\arraybackslash}m{#1}}

\begin{document}

\title{Seismocardiography for Emotion Recognition:\\A Study on EmoWear with Insights from DEAP}

\author{
  Mohammad~Hasan~Rahmani\orcidlink{0000-0002-0241-8270},
  Rafael~Berkvens\orcidlink{0000-0003-0064-5020},~\IEEEmembership{Member,~IEEE,}
  and~Maarten~Weyn\orcidlink{0000-0003-1152-6617},~\IEEEmembership{Member,~IEEE}%
  \thanks{This is the author’s version of the article that has been published in final form at IEEE Transactions on Affective Computing. The final version is available at: \href{https://doi.org/10.1109/TAFFC.2025.3575281}{10.1109/TAFFC.2025.3575281}.}
  \thanks{Mohammad Hasan Rahmani, Rafael Berkvens, and Maarten Weyn are with University of Antwerp - imec, IDLab - Faculty of Applied Engineering, Sint-Pietersvliet 7, 2000 Antwerp, Belgium (e-mail: mohammad.rahmani@uantwerpen.be; rafael.berkvens@uantwerpen.be; maarten.weyn@uantwerpen.be).}
}

\markboth{Preprint}%
{Rahmani \MakeLowercase{\textit{et al.}}: [Seismocardiography for Emotion Recognition: A Study on EmoWear with Insights from DEAP]}

\maketitle

\begin{abstract}
Emotions have a profound impact on our daily lives, influencing our thoughts, behaviors, and interactions, but also our physiological reactions.
Recent advances in wearable technology have facilitated studying emotions through cardio-respiratory signals.
Accelerometers offer a non-invasive, convenient, and cost-effective method for capturing heart- and pulmonary-induced vibrations on the chest wall, specifically \Ac{scg} and \ac{adr}.
Their affordability, wide availability, and ability to provide rich contextual data make accelerometers ideal for everyday use.
While accelerometers have been used as part of broader modality fusions for \ac{er}, 
their stand-alone potential via \ac{scg} and \ac{adr} remains unexplored.
Bridging this gap could significantly help the embedding of \ac{er} into real-world applications, minimizing the hardware, and increasing contextual integration potentials.
To address this gap,
we introduce \ac{scg} and \ac{adr} as novel modalities for \ac{er} and evaluate their performance using the EmoWear dataset.
First, we replicate the single-trial emotion classification pipeline from the DEAP dataset study, achieving similar results.
Then we use our validated pipeline to train models that predict affective valence-arousal states using \ac{scg} and compare them against established cardiac signals, \ac{ecgy} and \ac{bvp}.
Results show that \ac{scg} is a viable modality for \ac{er}, achieving similar performance to \ac{ecgy} and \ac{bvp}.
By combining \ac{adr} with \ac{scg}, we achieved a working \ac{er} framework that only requires a single chest-worn accelerometer.
These findings pave the way for integrating \ac{er} into real-world, enabling seamless affective computing in everyday life.
\acresetall
\end{abstract}


\section{Introduction}

\IEEEPARstart{E}{motion} recognition enhances human-computer inter\-action by enabling systems to understand and respond to users' emotional states~\cite{Wang2024}.
It has been studied and improved under the broader field of affective computing, with applications spanning healthcare~\cite{Hasnul2021}, entertainment~\cite{Paraschos2022}, education~\cite{haleem2023}, and marketing~\cite{Renjith2020}.
\textit{Emotions} are part of the broader category of \textit{affective states}, although the terms have been used interchangeably in the literature~\cite{Saganowski2022}, including in this paper.
In a reciprocal relationship, affective states influence physiological responses of the \ac{ans}, such as heart rate, breathing, and skin conductance~\cite{Kreibig2010a,Wac2014}.
Therefore, measuring physiological signals, including \ac{ecgy}, \ac{bvp}, \ac{rsp}, \ac{eda}, and \ac{skt}, provides indirect insights into the affective states, and can be used for \ac{er}~\cite{Saganowski2022}.
Recent advancements in wearable sensor technology have also facilitated unobtrusive and convenient \ac{er} research~\cite{Schmidt2019a}.

\begin{figure*}
\centering
\includegraphics[width=\linewidth]{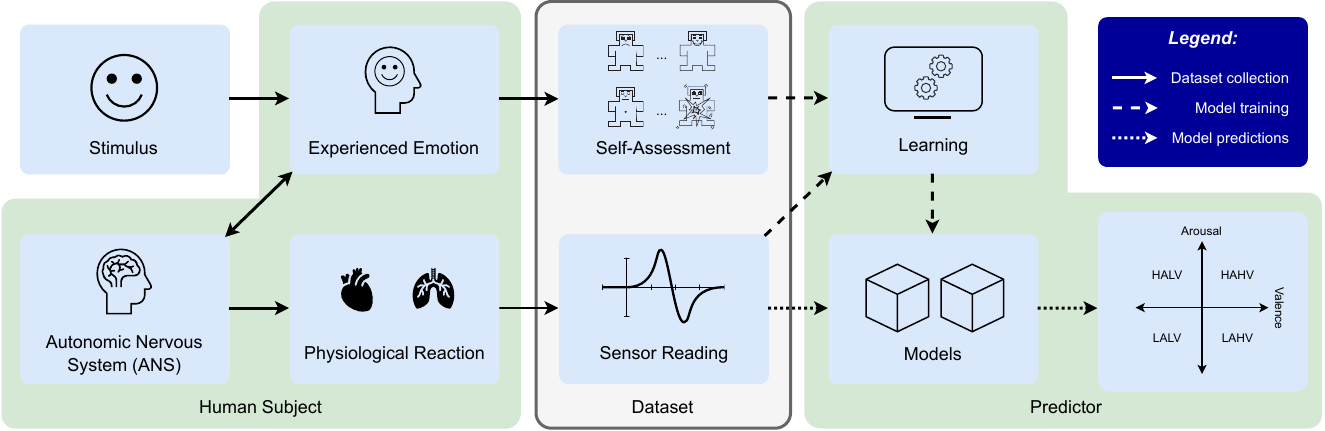}
\caption{Top-level overview of the hypothesis underlying this study: Emotion recognition research aims to map the objective representation of emotions in physiological signals to the subjective experiences of emotions, as captured by self-assessed emotional responses
(HAHV: High Arousal-High Valence, HALV: High Arousal-Low Valence, LALV: Low Arousal-Low Valence, LAHV: Low Arousal-High Valence).
\label{fig:hypothesis}}
\end{figure*}

Wearable-based \ac{er} research relies on the assumed link between the subjective emotional experiences and their objective representation in physiological signals~\cite{Saganowski2022}.
Emotional models try to explain these subjective experiences in structured and quantifiable terms.
They are often categorized in two broad types: \textit{discrete} and \textit{dimensional} models~\cite{Wang2022-slr}.
Discrete models represent emotions using distinct categories, such as happiness, sadness, anger, and fear, with Ekman's basic emotions model~\cite{Ekman1979} being a well-known example.
Dimensional models, on the other hand, represent emotions in a continuous space, typically using two or three dimensions.
Russell's circumplex model of affect~\cite{Russell1980} is a widely used dimensional model that represents emotions in a two-dimensional space: valence (pleasant vs. unpleasant) and arousal (activation vs. deactivation).
The limbic system, plays a key role in processing emotions and influencing the \ac{ans} through its connections with the hypothalamus~\cite{Martin2012}.
External stimuli can elicit emotional responses in the limbic system, which in turn manifest as physiological changes in the \ac{ans}, such as variations in breathing and heart rate~\cite{Martin2012, Appelhans2006}.
Researchers aim to develop models that map the objective representation of emotions in physiological signals to the subjective experiences of emotions, often measured using self-report or expert-assessed quantitative scales, such as the \ac{sam}~\cite{Bradley1994}.
Figure~\ref{fig:hypothesis} illustrates the top-level overview of the explained hypothesis, on which this study is based.

Currently, cardiac signals used for \ac{er} include \ac{ecgy} and \ac{bvp}~\cite{Hsu2020,Bugnon2020}.
Both of these signals measure the heart’s rhythmic activity, which is reflected in \ac{hrv} information.
\Ac{hrv} is modulated by the \ac{ans} in response to emotional stimuli~\cite{Appelhans2006}.
\Ac{scg}, which measures heart-originated vibrations on the chest wall, also carries \ac{hrv} information; however, its potential for \ac{er} remains unexplored~\cite{emowear}.
\Ac{scg} is a non-invasive method for sensing cardiac mechanical activity, typically captured through chest-worn accelerometers~\cite{Inan2015}.
It is often used for cardiac health monitoring through extracting the timing information~\cite{Rai2021}.

\Ac{scg} can be measured using a chest-worn accelerometer, offering a convenient and cost-effective method for capturing heart-induced vibrations on the chest wall.
In the field of \ac{er}, \ac{scg} is expected to provide \ac{hrv} information similar to \ac{ecgy} and \ac{bvp}, with the added advantage of being measured using the same accelerometer used for various other applications.
The way \ac{scg} is sensed, gives it a unique advantage over \ac{ecgy} and \ac{bvp} in terms of versatility and cost-effectiveness.
Accelerometers are widely available and can provide rich contextual data, making them ideal for everyday use.
The same device can support multiple applications, such as activity recognition, fall detection, and gait analysis.
In our previous work, we surveyed these applications and categorized their use cases across seven different domains~\cite{Rahmani2021}.
Another signal that can be inferred from the same chest-worn accelerometer is the \ac{adr}, which reflects the chest wall's movement due to respiration~\cite{Bricout2019}.
By combining \ac{scg} with \ac{adr}, we can infer both cardiac and respiratory signals from a single chest-worn accelerometer, offering a more comprehensive view of the physiological responses to emotions.

Combining \ac{scg} with \ac{adr} for \ac{er} is a practically interesting approach, as both can be captured using the same chest-worn accelerometer.
Theoretically, heart and respiration signals are tightly interconnected through the autonomic nervous system.
Heart activity is influenced by both sympathetic and parasympathetic nervous systems, which respectively accelerate or decelerate heart rate.
Respiratory rhythm is also controlled by autonomic inputs but can be modulated by conscious control.
During an emotional event, sympathetic arousal (as in fear or excitement) will typically raise heart rate and blood pressure, and also increase respiration rate and tidal volume~\cite{Critchley2012,Levenson1992A}.
Parasympathetic activity, on the other hand, promotes calm states, slowing the heart and stabilizing breathing.
By monitoring both signals, an \ac{er} system can capture these concurrent changes.

Several datasets have been developed to facilitate \ac{er} research using physiological signals~\cite{deap,emognition,mahnob-hci,case,popane,decaf,ascertain,biraffe,kemophone}
(for a comprehensive comparison of the existing datasets, we refer the reader to~\citetal{Kang}{kemophone}).
The EmoWear dataset~\cite{emowear} is the only one to provide chest-worn accelerometer data validated for both \ac{scg} and \ac{adr} purposes.
We collected and published this dataset to enable exploring \ac{scg} and \ac{adr} for \ac{er}.
EmoWear builds upon DEAP~\cite{deap} with the added elements of user mobility and context change.
Therefore, in terms of experimental setup and data collection process, DEAP is the closest dataset to EmoWear.
Both datasets are based on Russell’s dimensional circumplex model of affect~\cite{Russell1980}.
They both use video stimuli and collect self-assessed emotional responses using the \ac{sam}~\cite{Bradley1994} scale.
EmoWear uses stimuli from the same pool as DEAP (originally 40 stimuli, with 38 used in EmoWear). 
Both datasets provide cardiac and respiratory signal recordings, along with some additional peripheral modalities such as \ac{eda} and \ac{skt}.
Essentially, EmoWear has all the peripheral modalities that 
DEAP has, except for \ac{emgy} and \ac{eogy}, but plus \ac{scg} and \ac{adr}.
The similarities between the two datasets make DEAP an ideal benchmark for validating the EmoWear dataset in \ac{er} research, as it offers the highest resemblance when replicated.

This study evaluates \ac{scg} as a novel modality for \ac{er}, addressing a gap in the existing literature.
Moreover, we explore the potential of using a single chest-worn accelerometer for \ac{er} by combining \ac{scg} with \ac{adr}.
Motion was used as a modality for \ac{er} in the past~\cite{Schmidt2018,Hashmi2020}, but only in the context of measuring wrist or body movements.
This is the first study to investigate \ac{scg} as \iac{hrv}-providing alternative for \ac{er}.
Our approach aims to pave the way for integrating wearable-based \ac{er} in real-world applications.
Although our experiments are conducted on the EmoWear dataset, we also use the benchmark DEAP dataset to establish and strengthen our methodology and findings.
Our main research questions are:
1) Can \ac{scg} be effectively used for \ac{er}?
2) How do results from \ac{scg} compare to the conventional cardiac signals, \ac{ecgy} and \ac{bvp}?, and
3) How does a single chest-worn accelerometer perform for \ac{er} when combining \ac{scg} with \ac{adr}?

To address our research questions, we base our methodology on established best practices in the field.
A review of the existing literature provides both the necessary background and reference results for comparison.
However, this review revealed several methodological limitations in the current literature that could affect the interpretability and generalizability of the reported findings.
We aim to also address these limitations in our study and base our work on a robust and reliable methodology which is also comparable to existing results.
Following contributions are made in this paper:
\begin{itemize}
  \item A critical review of existing methodologies and reported results on the use of peripheral physiological signals (\ac{bvp} and \ac{rsp}, especially) from the DEAP dataset for \ac{er}.
  \item Replicating the DEAP dataset's single-trial emotion classification results, for reaffirming their findings and validating our analytical pipeline.
  \item Introduction of \ac{scg} as a novel gateway for \ac{er} and comparison of its performance against established cardiac signals.
  \item Examining the potential of a single chest-worn accelerometer for \ac{er} by combining \ac{scg} with \ac{adr}. 
  \item Setting the first benchmarks on the EmoWear dataset for \ac{er} research.
\end{itemize}

The rest of the paper is organized as follows:
Section~\ref{sec:related-work} reviews existing \ac{er} research on peripheral physiological signals from the DEAP dataset.
Section~\ref{sec:methodology} presents the methodology used in this study, including the datasets, signal processing, feature extraction, and machine learning models.
Section~\ref{sec:results} reports the experimental results, and Section~\ref{sec:discussion} discusses the findings.
Finally, Section~\ref{sec:conclusion} concludes the paper with a summary of the contributions and implications of the study.

%%%%%%%%%%%%%%%%%%%%%%%%%%%%%%%%%%%%%%%%%%%%%%%%%%%%%%%
\section{Related Work\label{sec:related-work}}

Review~\cite{Saganowski2022} distinguishes between \textit{classical feature-based} models and the \textit{end-to-end} deep learning models for \ac{er} using physiological signals from wearable sensors.
The former approach relies heavily on domain expertise for feature engineering, requiring the extraction and selection of meaningful physiological signal features before classification.
In contrast, the latter processes raw signals directly, automatically capturing temporal and spatial patterns inherent in physiological data, and learning the features that are most relevant for the task.
The end-to-end approach is a more recent development, used by only a minority of 12\,\% of their reviewed studies~\cite{Saganowski2022}.
Although this method shows promise by bypassing the limitations of traditional feature engineering, its limited use in wearable-based field studies may be attributed to the limited dataset sizes and computational resources.

More recent studies have addressed the limitations associated with the dataset size by using self-supervised representation learning~\cite{Sarkar2022,Wu2024} and transfer learning~\cite{Dresvyanskiy2022,Yen2022}.
In self-supervised learning the model is trained using automatically generated labels from the input data itself. 
It can be used to learn meaningful representations of specific signals (\textit{e.g.,}~\ac{ecgy}) on a large dataset to be used for a specific downstream task on a smaller dataset.
Transfer learning, on the other hand, is a technique where a model trained on one task is used for another related task.
Both techniques help in undermining the limitations of the dataset, and have shown promising results in \ac{er} research.

While these techniques show promise, our current study uses the traditional feature-based approach as we focus on providing initial insights into the potential of \ac{scg} for \ac{er} and the EmoWear dataset~\cite{emowear}.
EmoWear is a recently published dataset that offers the unique opportunity to explore the potential of \ac{scg} and \ac{adr} for \ac{er}.
It shares methodological similarities (see Section~\ref{sec:dataset}) with the benchmark DEAP dataset~\cite{deap}, which has been extensively used for \ac{er} research since its release in 2012.
In both datasets, video stimuli are presented to elicit emotions, concurrent with the recording of physiological signals and self-assessed emotional responses.

With the aim of learning from past work and identifying best practices for the EmoWear dataset, this section reviews existing \ac{er} research on peripheral physiological signals from the DEAP dataset.
This way we establish the rationale for our methodological approach as well as a ground to compare our results against.
Although the DEAP dataset contains a range of physiological recordings, our review focuses on studies that predict emotions using the \ac{bvp} and \ac{rsp} modalities, as these align with the scope of our current study.
Table~\ref{tab:related-work} provides a list of the related work that report findings on these specific modalities.
It shows a variety of combinations of methods used to measure the predictive power of different peripheral physiological signals for \ac{er}.
Different combinations of modalities, classifiers, data split approaches, performance metrics, and statistical analysis of the results, have been used to classify emotions and report findings.

Before diving into the details of these studies, it is important to note that the reviewed aspects are not exhaustive.
There are many other detail aspects differentiating the studies, that are not covered by the table, such as the preprocessing steps, feature extraction methods, and the hyperparameters of the classifiers.
For a more comprehensive analysis of the associated methods for \ac{er}, we refer the readers to the original cited references and plenty of review papers on the topic~\cite{Saganowski2022,Zhang2020-review,Schmidt2019a,Wijasena2021}.

A deeper look at the studies in Table~\ref{tab:related-work} reveal certain methodological limitations that potentially impact the interpretability and generalizability of the reported results.
We investigate these limitations across two main pillars: ``performance measurement and reporting'', and ``data splitting and validation''.
In what follows, we provide an overview of these limitations and highlight the key methodological gaps we found.
At the end of this section, we conclude with a summary of the identified gaps and the need for a robust and reliable methodology to address them.

\begin{table*}[!t]
  \caption{Overview of the related work that report on BVP and RSP modalities of the DEAP dataset. Several combinations of the methods are used to develop and evaluate the models.
  PS: Per Subject,
  MS: Mixed Subjects,
  SS: Separate Subjects,
  CV: Cross Validation,
  LOSO: Leave-One-Subject-Out,
  LOVO: Leave-One-Video-Out,
  A: Accuracy,
  $TP$: True Positive,
  $TN$: True Negative.
  ma: macro averaged,
  sc: single class,
  -: not reported,
  ?: not explicitly reported.
  \label{tab:related-work}}
  \centering
  \begin{tabular}{L{2.7cm}L{3.4cm}L{1.7cm}L{4.0cm}L{1.7cm}L{1.4cm}}
  \hline
  \textbf{Reference}
  & \textbf{Modalities}
  & \textbf{Classifier}
  & \textbf{Data split}
  & \textbf{Performance Metrics}
  & \textbf{Statistical Test}
  \\\hline
  
  \citetal{Koelstra}{deap} (Original DEAP paper)
  & BVP, RSP, EDA, SKT, EMG, EOG, EEG
  & NB
  & LOVO CV, PS
  & A \par $F1_{ma}$
  & one-sample
  \textit{t}-test
  \\\hline

  \citetal{Godin}{Godin2015}
  & BVP, RSP, EDA, SKT, EMG, EOG
  & NB
  & LOSO CV
  & A
  & -
  \\\hline
  \citetal{Chen}{Chen2016}
  & BVP, RSP, EDA, SKT, EMG, EOG, EEG
  & SVM
  & 10-fold CV, MS
  & A \par $F1_{ma}$
  & \textit{t}-test
  \\\hline
  \citetal{Zhang}{Zhang2017}
  & RSP
  & LR
  & 80\% training, 20\% testing, MS
  & A
  & -
  \\\hline
  \citetal{Yin}{Yin2017}
  & BVP, RSP, EDA, SKT, EMG, EOG, EEG
  & k-NN, LR, MESAE, NB, SVM
  & 10-fold CV, PS
  & A \par $F1_{sc}$
  & paired \textit{t}-test
  \\\hline
  \citetal{Ayata}{Ayata2018}
  & BVP, EDA
  & DT, RF, k-NN, SVM
  & 10-fold CV, MS
  & A
  & -
  \\\hline
  \citetal{Choi}{Choi2018}
  & BVP, EDA, EEG
  & LSTM
  & 80\% training, 20\% testing, SS (except for 1)
  & \raisebox{-3pt}{$\frac{TP_{sc?}}{TN_{sc?}}$}
  & -
  \\\hline
  \citetal{Lee}{FastLee2019} (2019)
  & BVP
  & CNN
  & 80\% training, 20\% testing, MS
  & A
  & -
  \\\hline
  \citetal{Lee}{Lee2020hc} (2020)
  & BVP
  & CNN
  & 80\% training, 20\% testing, MS (selected 20 out of 32)
  & A
  & -
  \\\hline
  \citetal{Wu}{Wu2020}
  & BVP, RSP, EDA, SKT, EMG, EOG, EEG
  & CNN, LSTM
  & -
  & A
  & -
  \\\hline
  \citetal{Zhu}{Zhu2020}
  & BVP, RSP, EDA, SKT, EMG, EOG, EEG, Face Video
  & CNN, MLP
  & 80\% training, 20\% testing, MS? (selected 18 out of 32)
  & A
  & -
  \\\hline
  \citetal{Elalamy}{Elalamy2021}
  & BVP, EDA
  & LR
  & LOSO CV
  & A \par $F1_{sc?}$
  & paired Wilcoxon test
  \\\hline
  \citetal{Kang}{Kang2022}
  & BVP, EDA
  & CAE
  & 64\% training, 16\% validation, 20\% testing, MS
  & A \par $F1_{sc?}$
  & -
  \\\hline
  \citetal{Pidgeon}{Pidgeon2022}
  & BVP, RSP, EDA, SKT
  & CNN
  & Stratified 10-fold CV, MS
  & A \par $F1_{sc?}$
  & -
  \\\hline
  \citetal{Siddharth}{Siddharth2022}
  & BVP, EDA, Face Video
  & -
  & LOSO CV
  & A \par $F1_{sc?}$
  & two-sample \textit{t}-test
  \\\hline
  \citetal{Shubha}{Shubha2023}
  & BVP, EDA
  & DT, RF, SVM, LR, AdaBoost
  & -
  & A \par $F1_{sc?}$
  & -
  \\\hline
  \citetal{Gohumpu}{Gohumpu2023}
  & BVP, EDA, SKT
  & DT, k-NN, SVM
  & 80\% training, 20\% testing, MS? + 2-fold CV
  & A \par $F1_{sc?}$
  & -
  \\\hline
  \citetal{Bamonte}{Bamonte2024}
  & BVP, EDA
  & k-NN, DT, RF, SVM, GBM, CNN
  & Stratified shuffle split PS (80\% training, 20\% testing repeated 10 times PS)
  & A \par $F1_{sc?}$
  & -
  \\\hline

  \end{tabular}

  \begin{minipage}{\linewidth}
  \vspace{0.2cm}
  \footnotesize{
  Abbreviations used in the table:
  BVP: Blood Volume Pulse, 
  RSP: Respiration, 
  EDA: Electrodermal Activity,
  SKT: Skin Temperature,
  EMG: Electromyography,
  EOG: Electrooculography,
  EEG: Electroencephalography,
  NB: Na{\"i}ve Bayes,
  SVM: Support Vector Machine,
  LR: Logistic Regression,
  DT: Decision Tree,
  RF: Random Forest,
  k-NN: k-Nearest Neighbors,
  LSTM: Long Short-Term Memory,
  CNN: Convolutional Neural Network,
  MLP: Multi-Layer Perceptron,
  CAE: Convolutional Autoencoder,
  MESAE: Multiple-fusion-layer based Ensemble classifier of Stacked Autoencoder (see~\cite{Yin2017}),
  GBM: Gradient Boosting Machine.
  }
  \end{minipage}
\end{table*}

\subsection{Performance Measurement and Reporting}

We identified several shortcomings in the performance measurement and reporting
of the reviewed studies, which we categorized in the following three areas:

\subsubsection{Accuracy and the Class Imbalance}

When the emotional valence and arousal in the DEAP dataset are binarized, a high class imbalance emerges~\cite{deap}.
Such imbalance can lead to biased classifiers that favor the majority class~\cite{Saganowski2022}, an effect that can be hidden when accuracy is the sole performance metric for evaluation.
Reporting robust performance metrics that are less sensitive to class imbalance, such as the $F1$ score, can help mitigate this issue~\cite{deap}.
Among the 18 reviewed studies, 7 (39\,\%) reported the accuracy as their sole performance metric, 1 (5\,\%) reported True Positive to True Negative ratio, and 10 (56\,\%) reported the $F1$ score along with the accuracy.
 
Some studies have used stratified data splitting techniques which ensures that the class distribution is preserved in the training and testing sets.
Although stratification may improve evaluation, the model still sees imbalanced data during its training and may still learn the biased patterns.
As a result, class stratification does not eliminate the need for robust performance metrics that are less sensitive to class imbalance.
Among the reviewed studies, 2 (11\,\%) used stratified data splitting techniques, both commonly reporting the $F1$ score as a performance metric.

\subsubsection{F1 Scoring Approach}

For imbalanced datasets, the $F1$ score is considered a more reliable performance metric than accuracy~\cite{deap}.
The $F1$ score is a harmonic mean of the precision and recall, which is designed to consider the trade-off between the two.
Both precision and recall are calculated based on the True Positive ($TP$), False Positive ($FP$), and False Negative ($FN$) counts in the classifier predictions.
The first letter ($T$ or $F$) in the abbreviations indicates the correctness of the prediction, and the second letter ($P$ or $N$) indicates the actual class of the sample.

When calculating the $F1$ score, it is necessary to define the class of interest upfront.
This is the class for which $TP$, $FP$, and $FN$ will be counted.
Defining the class of interest also implies that the $F1$ score is essentially a single-class performance metric, and the reported $F1$ score should be interpreted in the context of the pre-defined class.
Therefore, the following approaches can be considered when reporting the $F1$ score:
1) reporting $F1$ for the class of interest only, or
2) reporting $F1$ for all classes separately, or
3) reporting an average $F1$ over all classes.

For the classification of ``high'' vs. ``low'' emotions in the DEAP dataset, classes emerge from binarization of the continuous \ac{sam}~\cite{Bradley1994} that measure the emotional dimensions (pleasant vs. unpleasant for valence, and activation vs. deactivation for arousal).
In real-world applications, recognizing both high and low states is equally important as they represent the four quadrants of the valence-arousal space~\cite{Russell1980,Mehrabian1974}.
In such scenario, reporting a single-class $F1$ score is hard to interpret~\cite{Canbek2023,Jeni2013}, as it 
1) provides an incomplete picture of the classifier's performance, and
2) may lead to biased interpretations, especially if the chosen class is the majority class.
As a result, the $F1$ score should either be reported separately per class, or as an average $F1$ score over both classes.
The averaging can be weighted by the class distribution (weighted averaging), unweighted (macro averaging), or considered in the counting of the $TP$, $FP$, and $FN$ (micro averaging).
In the reviewed studies and among the 10 studies that reported the $F1$ score, 
7 (70\,\%) did not specify the class for which the $F1$ score was reported,
1 (10\,\%) reported the $F1$ score only for the ``low'' class, and
2 (20\,\%) reported the $F1$ score using the macro averaging approach.
With 16 (89\,\%) of the reviewed studies that aim to classify emotions using \ac{bvp} and \ac{rsp} modalities of the DEAP dataset not reporting any form of imbalance-tolerant metrics or not reporting them sufficiently, we identify a significant gap in performance reporting.
This gap may have led to severely inflated or biased conclusions in these papers.

\subsubsection{Statistical Significance Analysis}

The statistical signifi\-cance analysis help determining whether the observed differences in performance are due to the model's learning ability or random chance.
They strengthen the reliability of the reported performance and any conclusions drawn from it.
When the performance are averaged over groups of samples (\textit{e.g.,} folds in \ac{cv} or individual results in subject dependent classification), statistical significance tests can give an indication of the reliability of the reported average.
In case of comparing groups of results, these tests can help determine whether the observed differences are statistically significant~\cite{Field2002}.
Among the reviewed studies, 5 (28\,\%) conducted statistical significance tests, while 13 (72\,\%) did not report any statistical significance analysis.
Different forms of \textit{t}-tests, were the most commonly used statistical significance tests used by the reviewed studies.

\subsection{Data Splitting and Validation}

Data splitting strategies and validation techniques are crucial aspects of classification tasks.
Below we discuss the limitations observed in the reviewed studies from two perspectives: reliable generalization and transparency in selection criteria.

\subsubsection{Reliable Generalization}

An essential aspect of model evaluation is ensuring that results generalize across different data subsets.
The simplest way to achieve this is to split the dataset into two parts: a training set and a testing set.
The model is trained on the training set and evaluated on the testing set.
It is a simple and quick approach; however, it risks making the evaluation performance dependent on a very specific choice of data split.
Furthermore, if the test set is repeatedly used during model selection or hyperparameter tuning, the model would essentially learn the test set in an implicit way.
This may result in an overly optimistic estimate of performance that does not generalize to unseen data.
A common workaround is to include a validation set in the initial split.
The validation set is used for tuning and model selection, reserving the test set strictly for final evaluation. 
Without a validation set, it is hard to choose the best model or optimize hyperparameters without compromising the test set's role as an unbiased measure of generalization performance~\cite{Xu2018cv}.

With a single, fixed train-(validation)-test split, the performance metrics may not be representative of the entire dataset, as a different split could produce significantly different results.
A single, fixed split can lead to overfitting to the specific characteristics of that particular split.
\Ac{cv} techniques help to mitigate bias that can arise from relying on such a single, fixed data split.
\Ac{cv} evaluates the model on multiple subsets, providing a more comprehensive assessment of the model’s generalization performance.

Among the studies reviewed, 6 (33\,\%) did not use any form of \ac{cv} and instead reported performance based on a single data split.
Of these, 5 studies (28\,\%) did not dedicate a separate validation set, increasing the risk of overfitting to the test data.

\subsubsection{Transparency in Selection Criteria}

Another limitation observed in certain studies is the selective use of part of the dataset without providing a transparent selection criteria.
Selective sampling without explicit criteria risks introducing bias, especially if chosen samples share specific physiological traits or demographic characteristics.
Transparent and consistent selection methodologies are important to make sure that findings are representative of the broader dataset~\cite{Poldrack2017}.
Out of the reviewed studies, 2 (11\,\%) opted to use only a subset of the 32 subjects from the dataset but did not provide clear explanations for these selections.
Another 2 (11\,\%) do not share any information about their data splitting methodology, making their results hard to interpret.
 
In summary, the reviewed studies reveal several methodological gaps that can impact the reliability, interpretability, and generalizability of their findings.
Key issues in performance measurement and reporting include the lack of imbalance-tolerant metrics, imperfect $F1$ scoring approaches, and limited use of statistical significance testing. 
Moreover, the use of unreliable data splitting strategies and a lack of transparency in data selection further compromise the generalizability and trustworthiness of the findings in some of the reviewed studies.
These limitations highlight the need for robust and transparent methodologies to ensure that conclusions drawn from the results are both reliable and meaningful.
Future research must prioritize methodological rigor by systematically adopting best practices, including reliable data-splitting strategies, imbalance-tolerant metrics, proper statistical significance testing, and transparent reporting.
When using $F1$ score, especial care should be taken to report a complete picture of the way it was achieved, taking into account that, unlike accuracy, it is essentially a single-class metric.
Without these foundational elements, advancements in emotion recognition risk being built on fragile methodological grounds.
We urge the community to integrate these practices as standard, rather than optional, to improve the reliability, interpretability, and generalizability of findings in this domain.
In the next section, we present our method, designed to address these methodological gaps. 
We use a validated approach to deliver results that set reliable benchmarks on the EmoWear dataset. 
First, we validate the emotion classification pipeline used in the DEAP study on the DEAP dataset. 
Then, we apply the same pipeline to the EmoWear dataset.

\section{Methodology\label{sec:methodology}}

\begin{figure*}
\centering
\includegraphics[width=0.9\textwidth]{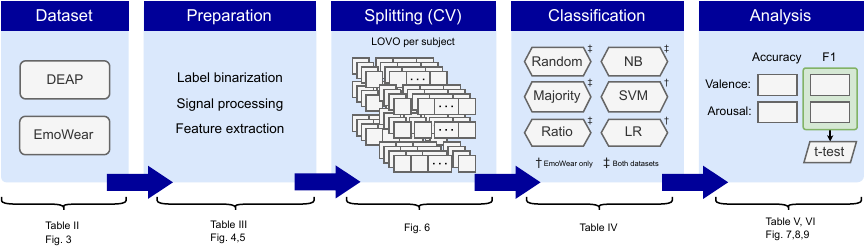}
\caption{Overview of the emotion classification pipeline used in this study. The pipeline is inspired by the single-trial emotion classification pipeline from the DEAP study, and is applied to both the DEAP and EmoWear datasets (CV: Cross Validation, LOVO: Leave-One-Video-Out, NB: Na{\"i}ve Bayes, SVM: Support Vector Machine, LR: Logistic Regression).
\label{fig:methodology-overview}}
\end{figure*}

Our study can be divided into two main parts: the replication of the DEAP dataset's emotion classification pipeline, and the application of this pipeline to the EmoWear dataset.
First, we replicated the single-trial emotion classification pipeline from the DEAP dataset and obtained similar results.
Next, we applied the same approach to the EmoWear dataset, which offers a new proxy signal, \ac{scg}, for \ac{er}.
By first confirming our implementation using the DEAP benchmark, we ensured that we were applying a validated implementation to the EmoWear dataset, thereby better supporting our findings and providing more meaningful insights into the performance of the new proxy signal, \ac{scg}, for \ac{er}.
Figure~\ref{fig:methodology-overview} provides an overview of the emotion classification pipeline used in this study.

\subsection{Dataset\label{sec:dataset}}

We used the DEAP dataset~\cite{deap} and the EmoWear dataset~\cite{emowear} for our study.
These two datasets share the stimulus videos and the \ac{sam} rating scales, but differ in the recording equipment and the subjects.
DEAP is a benchmark dataset for \ac{er}, recorded in a laboratory setting with non-portable sensors, while EmoWear is a recent dataset, recorded with portable wearable sensors.
Looking at the sensor modalities of our interest, DEAP offers \ac{bvp} for cardiac-, and \ac{rsp} for respiratory-sensing.
EmoWear offers \ac{bvp} and \ac{ecgy} for cardiac-, and \ac{rsp} for respiratory-sensing.
EmoWear also offers the chest-recorded vibration data through \ac{acc} which we use to infer both the \ac{scg} and the \ac{adr} signals for cardiac- and respiratory-sensing, respectively.
Table~\ref{tab:dataset-summary} provides a content summary of the two datasets.

\begin{table}[!t]
  \caption{Content summary of EmoWear and DEAP datasets.\\
  (F: Female, M: Male.)
  \label{tab:dataset-summary}}
  \centering
  \begin{tabular}{L{2.2cm}|L{2.5cm}|L{2.5cm}}
  \hline
  & \textbf{DEAP}~\cite{deap} & \textbf{EmoWear}~\cite{emowear, emowear-data-ua}\\
  \hline
  Participants & 32 (15\,F, 17\,M) & 48 (21\,F, 27\,M)\\
  \hline
  Recorded signals & EEG, EDA, BVP, RSP, SKT, EMG, EOG, Face Video & ACC (SCG \& ADR), GYRO, ECG, BVP, RSP, EDA, SKT\\
  \hline
  Signal recording equipment & Non-portable sensors & Portable wearable sensors\\
  \hline
  Number of stimuli per participant & 40 (10 HAHV, 10 LAHV, 10 LALV, 10 HALV) & 38 (10 HAHV, 9 LAHV, 10 LALV, 9 HALV)\\
  \hline
  Rating method & Self-assessment manikins (SAM) & Self-assessment manikins (SAM)\\
  \hline
  Rating scales & Arousal, Valence, Dominance, Liking, Familiarity & Arousal, Valence, Dominance, Liking, Familiarity\\
  \hline
  Stimulus presenter / mediator GUI & Presentation\,®~\cite{presentation-neurobs} & ColEmo~\cite{colemo}\\
  \hline
  Publication year & 2012 & 2024\\
  \hline
  \end{tabular}
\end{table}

\subsubsection*{Ethical Statement}

In this study, we are using the publicly available datasets in accordance with their respective terms of use.
The EmoWear dataset was collected by us in accordance with the ethical guidelines of the University of Antwerp, and following a positive ethical clearance decision.
Our use of the DEAP dataset is bounded by their end-user license agreement which is limited for research purposes.
We use the physiological signals present in the DEAP dataset from those subjects who have given their consent for the distribution of their physiological recordings.

\subsubsection*{Packages Used}

We used the preprocessed DEAP dataset package and the second phase (elicit-assess-walk cycles) of the EmoWear dataset's CSV package~\cite{emowear-data-ua}.
The preprocessed DEAP dataset offers a ready-to-use downsampled, filtered, and segmented version of the physiological signals, offered in Matlab and Python pickle formats.
We used the Python pickle format for our study.
The EmoWear dataset is recorded in two phases: the first phase is the vocal vibration recording, and the second phase involves the elicit-assess-walk cycles.
We used the second phase data which is the relevant phase for \ac{er} research.
The dataset is offered in three packages: the raw data package, the Matlab data package, and the CSV package.
The CSV package offers the synchronized and segmented data of the physiological signals and the self-assessed emotional responses, which we used for our study.

\subsubsection*{Subject Demographics}
The DEAP dataset contains 32 subjects (15 females and 17 males) aged between 19 and 37 years (mean=26.9), while the EmoWear dataset contains 48 subjects (21 females and 27 males) aged between 21 and 45 years (mean=29.3).
One subject in the DEAP dataset reported a history of migraine, while three subjects in the EmoWear dataset reported histories of ADHD\,-\,autism, panic attacks, and epilepsy.

\subsubsection*{Exclusion Criteria}

\begin{figure}
\centering
\includegraphics[width=\linewidth]{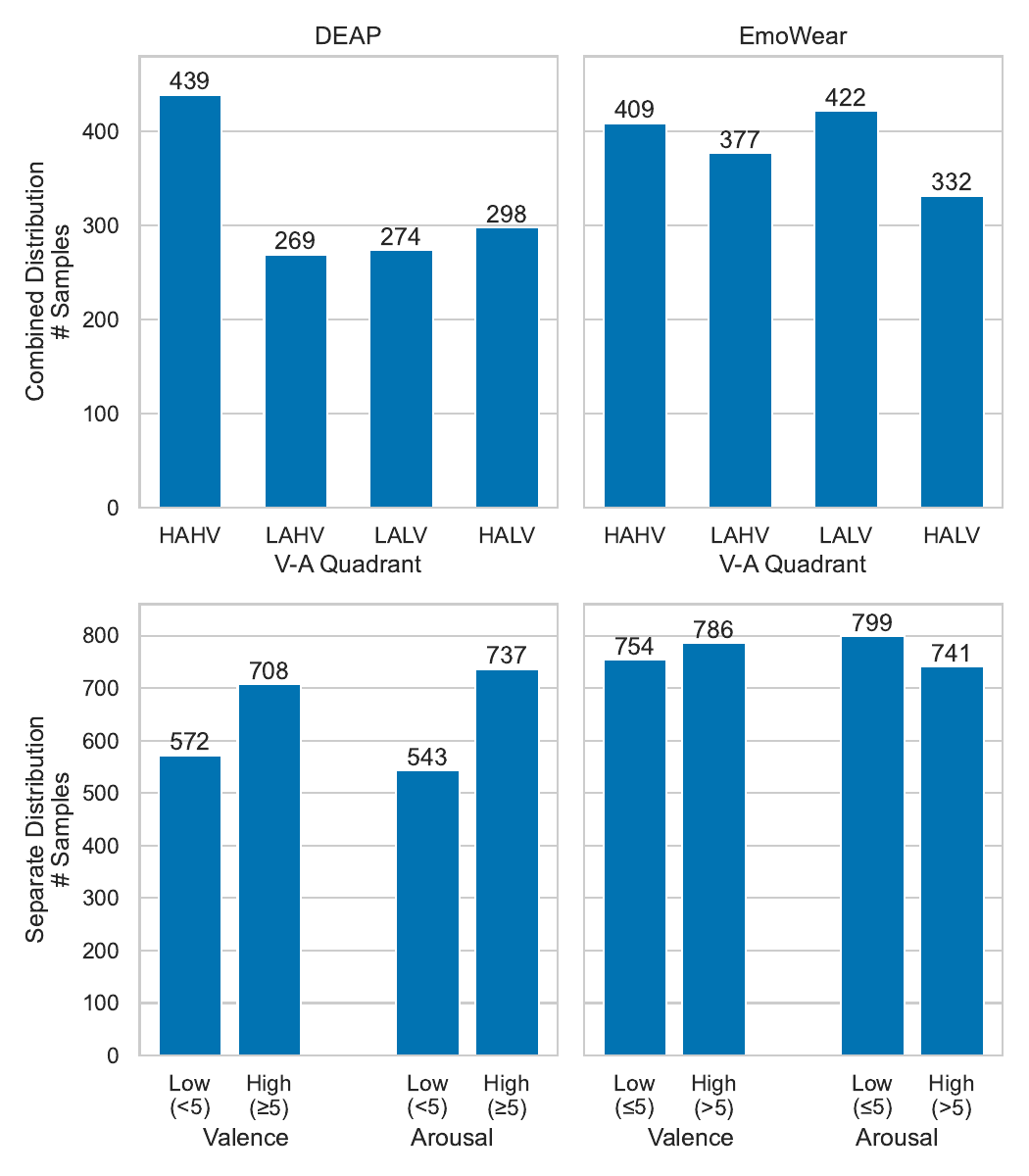}
\caption{Class distribution of the DEAP and EmoWear datasets, after applying the exclusion criteria.
\label{fig:data-dist}}
\end{figure}

We excluded subjects whose frontal chest accelerometer data was missing or corrupted (subjects 1, 3, 35 from the EmoWear dataset).
We also excluded subjects whose emotion ratings were severely imbalanced.
To this end, we considered an empirical threshold of 10\,\% for availability of high and low ratings for both binarized valence and arousal, below which the subject was excluded.
This criterion resulted in the exclusion of subjects 19, 21, and 32 from the EmoWear dataset.
Also, we excluded subject 34 from the EmoWear dataset due to the use of a different sensor unit for the chest accelerometer data.
The final number of subjects used in our study was 42 from the EmoWear dataset.
In addition to the mentioned subject exclusion criteria, we excluded single trials identified as faulty in the original EmoWear data description~\cite{emowear-data-ua}.
These faulty trials were identified based on severe disruptions during the experiments where:
1) participant was distracted by external factors during stimulus presentation (accidental objects falling), and
2) video stimulus was not presented correctly (sound issues).
For the DEAP dataset, no subjects or trials were excluded because none met any of the specified exclusion criteria.
Figure~\ref{fig:data-dist} shows the class distribution of the DEAP and EmoWear datasets after applying the exclusion criteria.

\subsection{Signal Processing}

Signal processing involved preprocessing the raw physio\-logical signals, and extracting features from them.
While all other signals have been directly recorded, the \ac{scg} and \ac{adr} signals are derived from the \ac{acc} signal.
Therefore, we first describe the signal processing steps for the \ac{scg} and \ac{adr} signals, separately in detail, followed by the preprocessing steps for the other signals.
All signal processing including preprocessing and feature extraction was done in Python~3.10.14 using the \textit{NumPy}~\cite{numpy}, \textit{pandas}~\cite{pandas-sw, pandas-paper}, and the \textit{NeuroKit2}~\cite{neurokit2} libraries.

\subsubsection*{Seismocardiography}
\begin{figure*}[!t]
\centering
\begin{tikzpicture}
    \node[anchor=south west,inner sep=0] (a) at (0,0) {\subfloat[]{\includegraphics[width=3.5in]{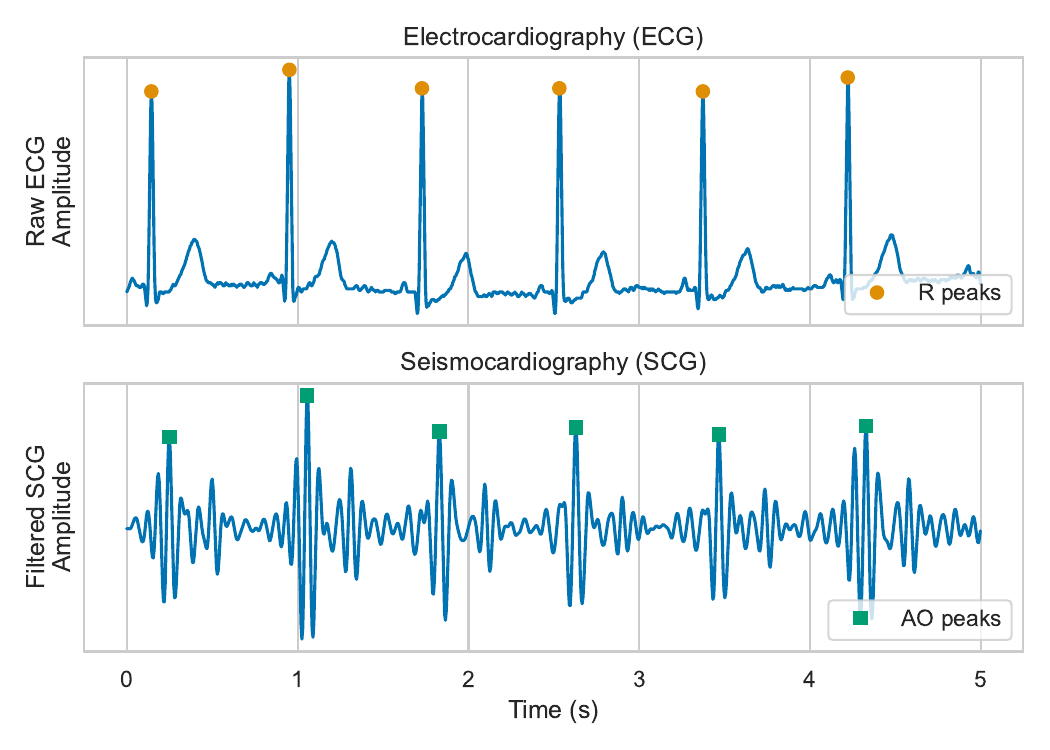}%
    \label{fig:ecg_scg_peaks}}};
    \node[anchor=south west,inner sep=0] (b) at (9,0) {\subfloat[]{\includegraphics[width=3.5in]{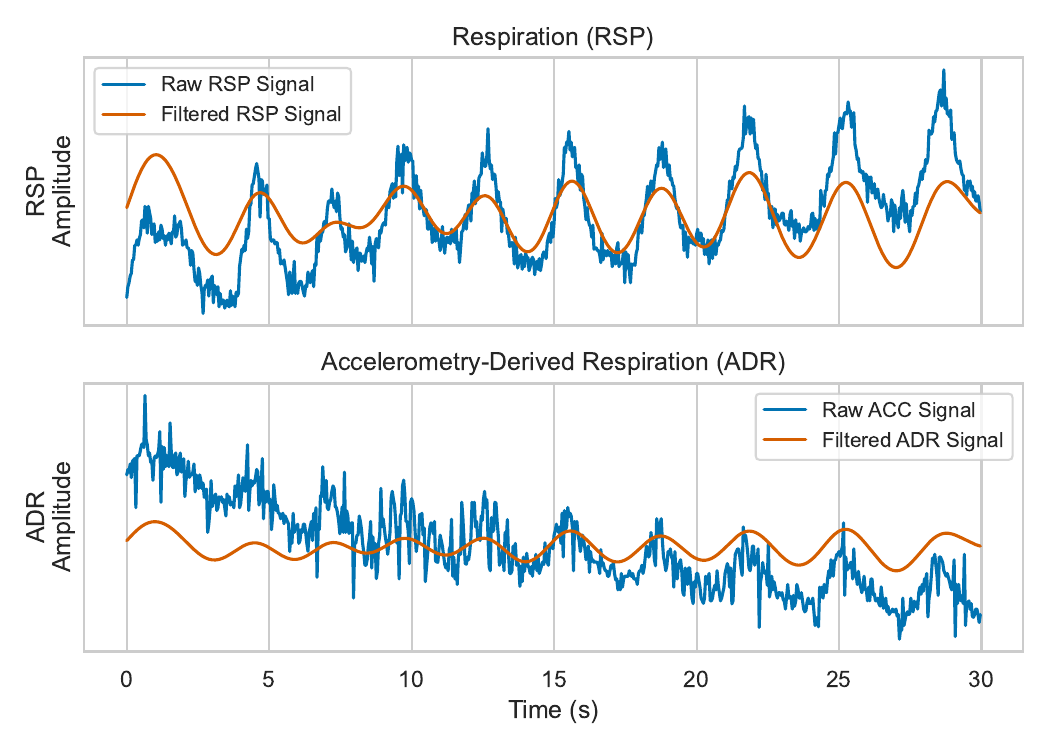}%
    \label{fig:rsp_adr_plots}}};

    \begin{scope}[x={(a.south east)},y={(a.north west)}]
        \node[draw=blue, thick, rectangle, anchor=south west] (ecg-scg) at (0.01,0.1) [minimum width=8.8cm, minimum height=3.35cm] {};

        \node[draw=blue, thick, rectangle, anchor=south west] (rsp-adr) at (1.02,0.1) [minimum width=8.8cm, minimum height=3.35cm] {};
      \end{scope}
      
      \node (guide) [text width=4cm,align=center,draw=blue] at (9.1,0.2) {\small Captured by Accelerometer};
      \draw[->,blue,thick] (ecg-scg) -- (guide);
      \draw[->,blue,thick] (rsp-adr) -- (guide);
\end{tikzpicture}
\caption{Example signals from the EmoWear dataset representing how chest-worn accelerometer data hold cardio-respiratory information. (a) Electrocardiography (ECG) and Seismocardiography (SCG) signals with detected R-peaks and AO-peaks. (b) Respiration (RSP) and Accelerometry-Derived Respiration (ADR) before and after filtering.
\label{fig:example_signals}}
\end{figure*}
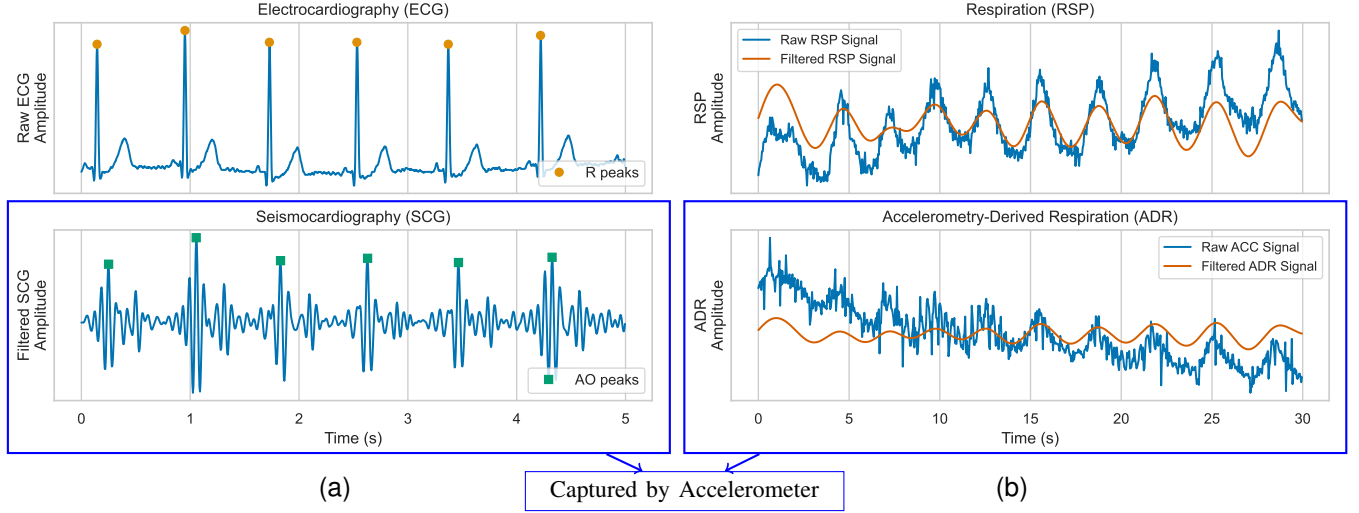

The so-called R-peaks of the \ac{ecgy} signal represent ventricular depolarization, an elec\-trical event that initiates the mechanical contraction of the left ventricle.
Once the pressure in the left ventricle exceeds the pressure in the aorta, the aortic valve opens~\cite{Pollock2017}, leading to \iac{ao} peak in the \ac{scg} signal.
The aortic valve opens shortly after the R-peak due to the time required for ventricular contraction to generate sufficient pressure to overcome aortic pressure~\cite{Dehkordi2019b, Vaini2015}.
This corresponding relationship makes the \ac{ao} peak our best alternative to the R-peak for cardiac activity sensing in the \ac{scg} signal.
Figure~\ref{fig:ecg_scg_peaks} shows an example of the \ac{ecgy} and \ac{scg} signals from the EmoWear dataset with marked R-peaks and AO-peaks.

We use the \ac{ao} peaks for our analysis of the \ac{hr} and \ac{hrv} indexes from the \ac{scg} signal.
We took the data of dorsoventral (z) axis from the frontal $ACC_3$ (LSM6DSOX) sensor of the EmoWear dataset.
Since the accelerometer data was sampled irregularly~\cite{emowear}, we interpolated the data to a regular 200\,Hz sampling rate, which we used as our \ac{scg} signal.

Next, we detected the \ac{ao} peaks in the \ac{scg} signal using the algorithm proposed by \citetal{Massaroni}{Massaroni2022}.
Our implementation involves an initial band-pass filtering with a pass-band of 10--20\,Hz, envelope detection using the Hilbert transform, a second band-pass filtering with a pass-band of 0.5--2\,Hz, and a final peak detection algorithm.
Figure~\ref{fig:ecg_scg_peaks} shows detected AO-peaks in an initially-filtered \ac{scg} signal.

\subsubsection*{Accelerometry-Derived Respiration}

Chest-sensed accel\-e\-ration signals also carry respiratory information~\cite{Rahmani2021}.
Many studies have validated \ac{adr} against the gold standard respiratory signals, such as spirometry~\cite{Lapi2014} and \ac{rip}~\cite{Bricout2019}.
We derive the \ac{adr} signal from the chest-sensed \ac{acc} signal by applying a band-pass filter of 0.15--0.35\,Hz to extract the respiratory frequency band, and then a baseline detrending (same methodology of BioSPPy toolbox~\cite{biosppy}).
Figure~\ref{fig:rsp_adr_plots} shows an example of the \ac{rsp} (using capacitive breathing sensor of Zephyr Bio\-Harness~3~\cite{emowear}) and \ac{adr} signals from the EmoWear dataset, before and after the mentioned filtering/detrending steps.

\subsubsection*{Peripheral Physiological Signals}

\begin{figure}[!t]
\centering
\includegraphics[width=\linewidth]{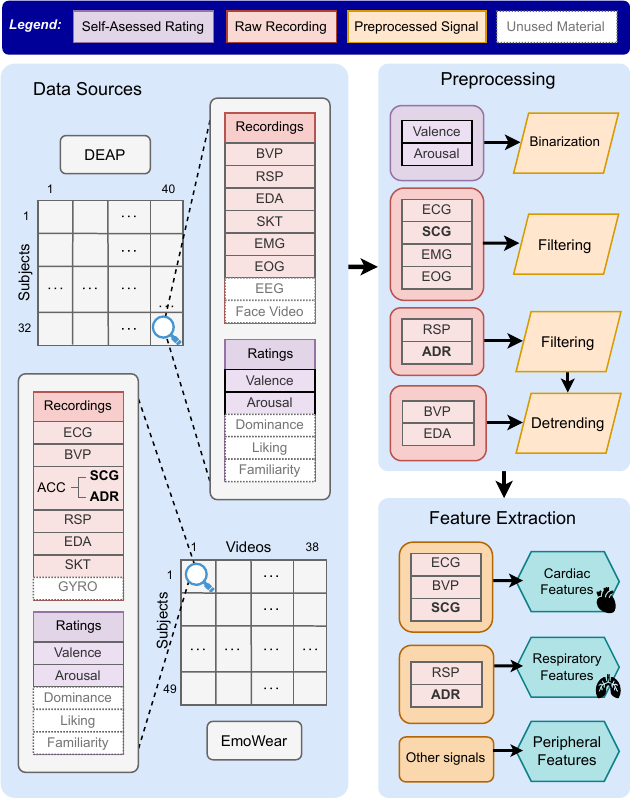}
\caption{Signal processing and feature extraction pipeline for different types of physiological signals in both DEAP and EmoWear datasets.
\label{fig:method_sig}}
\end{figure}

Figure~\ref{fig:method_sig} shows a summary of the data sources used in our study and the signal processing pipeline applied to them.
We used the \ac{bvp}, \ac{rsp}, \ac{eda}, \ac{emgy}, \ac{eogy}, and \ac{skt} signals from the DEAP dataset, and the \ac{ecgy}, \ac{bvp}, \ac{scg}, \ac{rsp}, \ac{adr}, \ac{eda}, and \ac{skt} signals from the EmoWear dataset.

Different signals in both datasets were preprocessed to remove noise and artifacts, and prepare them for feature extraction.
The \ac{ecgy} signal was band-pass filtered between 8--20\,Hz~\cite{Elgendi2010}.
The \ac{bvp} and \ac{eda} signals were detrended by subtracting a 256-point moving average from the signal~\cite{deap}.
Similar to the \ac{adr}, the \ac{rsp} signal was band-pass filtered between 0.15--0.35\,Hz~\cite{biosppy}.
Finally, the \ac{emgy} and \ac{eogy} signals were both band-pass filtered between 4--40\,Hz~\cite{deap}.
The \ac{skt} signal was not preprocessed.

\begin{table}
\caption{Summary of Extracted Features from Physiological Signals\label{tab:features}}
\centering
\begin{tabular}{L{0.8cm}|L{7.0cm}}
\hline
\textbf{Signal} & \textbf{Extracted Feature} \\
\hline
ECG, BVP, SCG
& Mean and standard deviation of inter-beat intervals (IBI), heart rate (HR), and IBI derivatives.
Frequency band power ratio (0.04--0.15\,Hz vs. 0.15--0.5\,Hz) for IBI.
Spectral power in IBI bands: 0.1--0.2\,Hz, 0.2--0.3\,Hz, 0.3--0.4\,Hz.
Spectral power in IBI derivatives: low (0.01--0.08\,Hz), medium (0.08--0.15\,Hz), and high (0.15--0.5\,Hz) bands.\\
\cline{2-2}
& \textit{Additionals$^*$:} CVNN, CVSD, HFn, HTI, IQRNN, LFn, LnHF, MCVNN, MadNN, MaxNN, MedianNN, MinNN, Prc20NN, Prc80NN, RMSSD, SDRMSSD, TINN, TP, pNN20, pNN50\\
\hline
RSP, ADR
& Mean and standard deviation of signal.
Mean of signal derivatives.
Breathing rate.
Mean and median of peak-to-peak intervals.
Logarithmic band energy difference (0.05--0.25\,Hz and 0.25--5\,Hz).
Spectral centroid of breathing rhythm.
Band power up to 2.4\,Hz.\\
\cline{2-2}
& \textit{Additionals$^*$:} CVSD, RMSSD, ApEn, CVBB, HF, LF, LFHF, MCVBB, MadBB, SD1, SDBB, SDSD, RVT\\
\hline
EDA
& Mean resistance and mean derivative values.
Mean decrease rate during decay periods.
Proportion of negative derivative samples.
Local minima count.
Mean rise time.
Band power up to 2.4\,Hz.
Zero-crossing rates for skin conductance slow, and very slow responses (SCSR and SCVSR). 
Mean peak magnitudes of SCSR and SCVSR. \\
\hline
SKT & Mean temperature, mean of derivatives, and spectral power in low-frequency bands (0--0.1\,Hz and 0.1--0.2\,Hz). \\
\hline
EMG & Signal energy, and basic statistical measures (mean, variance). \\
\hline
EOG & Same as EMG + eye blinking rate. \\
\hline
\end{tabular}
\begin{minipage}{\linewidth}
\vspace{0.2cm}
\footnotesize{$^*$For a full description of the terms of the additional features, visit the NeuroKit2 documentation~\cite{neurokit2-doc}.}
\end{minipage}
\end{table}

\subsubsection*{Feature Extraction\label{sec:features}}

Different hand-crafted features were extracted from the preprocessed signals based on the literature.
We extracted the set of features that were used by the DEAP dataset to enable replication of their results.
For the cardiac (\ac{ecgy}, \ac{bvp}, and \ac{scg}), and the respiratory (\ac{rsp} and \ac{adr}) signals, we also extracted additional features than those reported in the DEAP dataset based on more recent common practices in the literature.
The additional features were only extracted from the EmoWear dataset.
Table~\ref{tab:features} lists the features extracted from the different physiological signals.

\subsection{Machine Learning\label{sec:ml}}

\subsubsection*{Classification Tasks}
The circumplex model of affect~\cite{Russell1980} defines emotions in a two-dimensional space of valence and arousal.
This model assumes independence between the two dimensions~\cite{Saganowski2022}.
Based on this assumption, we considered two independent classification problems: the binary classification of either of the emotional dimensions - valence and arousal.
The binarization of the continuous \ac{sam} ratings into high and low values defines the classes.

\subsubsection*{Input Sets}
We fed the machine learning classifiers with the hand-crafted features explained in Section~\ref{sec:features}.
For the DEAP dataset, we used all the available peripheral physiological signals, to replicate their results (similar to the original DEAP study).
For the EmoWear dataset, we experimented with two types of input set combinations:
1) all-modality sets: one of the cardiac recordings (ECG\,/\,BVP\,/\,SCG) along with all other available peripheral signals (RSP,\,ADR,\,EDA,\,and\,SKT), and
2) cardio-respiratory sets: one of the cardiac recordings (ECG\,/\,BVP\,/\,SCG) along with one of the respiratory recordings (RSP\,/\,ADR).
In either case, we never combined ADR with ECG or BVP, as ADR is derived from the same sensor as SCG and is therefore only available when SCG is.
The all-modality sets provide a ground for comparison with the DEAP dataset using various cardiac sources, while the cardio-respiratory sets allow us to further investigate the potential of the SCG and ADR signals (essentially, a single chest-worn accelerometer) for \ac{er}.
Table~\ref{tab:inputs} summarizes the modality-combination input scenarios experimented in the study.

\begin{table*}[!t]
\caption{Summary of Classifier Input Scenarios Experimented in the Study\label{tab:inputs}}
\centering
\begin{tabular}{c|c|ccc|cc|cccc|c|c}
\hline
\multirow{2}{*}{\textbf{Input Category}} & \multirow{2}{*}{\textbf{Dataset}} & \multicolumn{3}{c|}{\textbf{Cardiac}} & \multicolumn{2}{c|}{\textbf{Respiratory}} & \multicolumn{4}{c|}{\textbf{Other Peripherals}} &  \multirow{2}{*}{\textbf{Input Scenario}} & \multirow{2}{*}{\textbf{Presented Results}}\\
\cline{3-11}
& & BVP & ECG & SCG & RSP & ADR & EDA & SKT & EMG & EOG & & \\
\hline
Replication 
& DEAP
& \checkmark % BVP
& % ECG
& % SCG
& \checkmark % RSP
& % ADR
& \checkmark % EDA
& \checkmark % SKT
& \checkmark % EMG
& \checkmark % EOG
& BVP\,+\,all 
& Table~\ref{tab:results-deap} \\\hline
\multirow{3}{2.1cm}{\centering All-Modality \par Sets}
& \multirow{7}{*}{EmoWear}
& \checkmark % BVP
& % ECG
& % SCG
& \checkmark % RSP
& % ADR
& \checkmark % EDA
& \checkmark % SKT
& % EMG
& % EOG
& BVP\,+\,all
& \multirow{3}{1.9cm}{\centering Table~\ref{tab:results-emowear}, Fig.~\ref{fig:boxplots}, Fig.~\ref{fig:f1_heatmap}} \\
& % EmoWear
& % BVP
& \checkmark % ECG
& % SCG
& \checkmark % RSP
& % ADR
& \checkmark % EDA
& \checkmark % SKT
& % EMG
& % EOG
& ECG\,+\,all &  \\
& % EmoWear
& % BVP
& % ECG
& \checkmark % SCG
& \checkmark % RSP
& \checkmark % ADR
& \checkmark % EDA
& \checkmark % SKT
& % EMG
& % EOG
& \textbf{SCG}\,+\,all &  \\\cline{1-1}\cline{3-13}
\multirow{4}{2.2cm}{\centering Cardio-Respiratory \par Sets}
& % EmoWear
& \checkmark % BVP
& % ECG
& % SCG
& \checkmark % RSP
& % ADR
& % EDA
& % SKT
& % EMG
& % EOG
& BVP\,+\,RSP
& \multirow{4}{1.9cm}{\centering Table~\ref{tab:results-emowear}, Fig.~\ref{fig:boxplots}, Fig.~\ref{fig:cardio-resp-compare}, Fig.~\ref{fig:f1_heatmap}} \\
& % EmoWear
& % BVP
& \checkmark % ECG
& % SCG
& \checkmark % RSP
& % ADR
& % EDA
& % SKT
& % EMG
& % EOG
& ECG\,+\,RSP & \\
& % EmoWear
& % BVP
& % ECG
& \checkmark % SCG
& \checkmark % RSP
& % ADR
& % EDA
& % SKT
& % EMG
& % EOG
& \textbf{SCG}\,+\,RSP & \\
% Chest-Worn Accelerometer
& % EmoWear
& % BVP
& % ECG
& \checkmark % SCG
& % RSP
& \checkmark % ADR
& % EDA
& % SKT
& % EMG
& % EOG
& \textbf{SCG\,+\,ADR} & \\
\hline
\end{tabular}
\end{table*}

\subsubsection*{Target Labels}
We binarized the self-assessed valence and arousal ratings into high and low values based on the mid-threshold of 5 in their 1--9 scales.
Such binarization divides the valence-arousal space into four quadrants: High Arousal-High Valence (HAHV), Low Arousal-High Valence (LAHV), Low Arousal-Low Valence (LALV), and High Arousal-Low Valence (HALV).
The dominance, liking, and familiarity dimensions were not used in our study.

\subsubsection*{Feature Selection}
We use Fisher's linear discriminant score to rank the features based on their discriminative power.
The Fisher score \textit{J} of a feature \textit{f} is calculated as:
\begin{equation}
  J(f) = \frac{|\mu_1 - \mu_2|}{\sigma_1^2 + \sigma_2^2}
\end{equation}
where $\mu_1$ and $\mu_2$ are the means, and $\sigma_1$ and $\sigma_2$ are the standard deviations of the feature \textit{f} for the two classes.
The feature was selected if its Fisher score was above the empirical threshold of 0.3.
For the EmoWear dataset, we combined this threshold with a pre-determined minimum ranked feature count of 15 to ensure that the feature set did not become too small.

\subsubsection*{Classifiers}
We used three classic classifiers: 
Gaussian \ac{nb}, \ac{svm}, and \ac{lr}.
The \ac{nb} classifier is a simple probabilistic classifier based on the Bayes theorem with strong independence assumptions between the features.
Despite the fact that the assumption of independence does not apply to our type of extracted features (Table~\ref{tab:features}), the \ac{nb} classifier has been successfully employed by similar studies~\cite{deap,Saganowski2022}.
However, our choice of this classifier was mainly based on:
1) replicating the DEAP dataset's analytical pipeline,
2) enabling a direct comparison of the results.

The choice of \ac{svm} and \ac{lr} classifiers was based on our initial investigation of different classifiers and hyperparameters.
We took k-Nearest Neighbors, Decision Trees, Random Forest, boosting classifiers (including Adaboost, Gradient Boosting, and XGBoost), and Neural Networks (including \ac{mlp}, \ac{cnn}, \ac{lstm}).
We experimented with different hyperparameters and optimized their performance using a grid search.
The goal was to maximize the macro-averaged F1 score over the high and low classes of the emotional dimensions in the entire EmoWear dataset.
The best performing classifiers were the \ac{svm} and \ac{lr} classifiers with L2 regularization, balanced class weights, and linear kernels.
Based on this initial investigation, we limited our scope to these classifiers, only varying the regularization strength parameter (usually known as \textit{C} parameter) for both classifiers, and the solver parameter for the \ac{lr} classifier.

\subsubsection*{Baseline Scores}
To establish a baseline for our study, we implemented three simple voting strategies: 
1) random voting, which assigns a class label randomly to each given sample; 
2) majority voting, which assigns the most frequently occurring class label from the subject's labels to each sample; and 
3) ratio voting, which assigns a class label randomly based on a probability distribution proportional to the ratio of class labels for the subject.
These strategies were used to compare the results with the expected values of analytically determined baseline scores.

\subsubsection*{Model Validation}
\begin{figure}[!t]
\centering
\includegraphics[width=\linewidth]{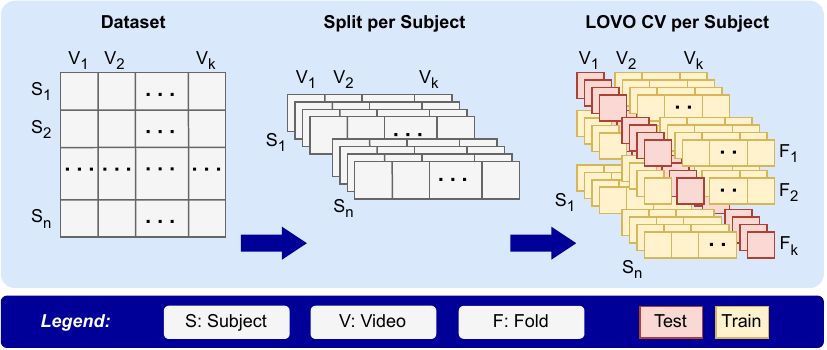}
\caption{Data splitting and Cross Validation (CV) scheme of the study. Leave-One-Video-Out (LOVO) CV was applied to every subject.
\label{fig:method_cv}}
\end{figure}

Figure~\ref{fig:method_cv} shows the \ac{cv} scheme used in our study.
We develop subject-dependent classification models for both datasets, where one video from a subject at a time is used for testing while the rest are used for training.
This splitting is repeated for each video of the subject using a Leave-One-Video-Out (LOVO) \ac{cv} scheme.
Each subject receives an accuracy and a macro-averaged $F1$ score based on the confusion matrix of the subject's predictions.
The final performance metrics are the averages of these subject-wise scores, calculated separately for each emotional dimension.
This approach complies with the DEAP performance evaluation methodology, and allows for a direct comparison of the results.
To assess statistical significance, a one-sample \textit{t}-test is performed, comparing the average $F1$ score to the best value set by the baseline voting strategies.

\section{Results\label{sec:results}}

\subsection{Replication of DEAP Results~\label{sec:results-deap}}

As first step of our study, we replicated the single-trial emotion classification pipeline from the DEAP dataset.
We used the preprocessed DEAP dataset package and focused on classifying the valence and arousal dimensions using peripheral physiological signals.
The list of peripheral signals was identical to those used in the original DEAP paper: \ac{bvp}, \ac{rsp}, \ac{eda}, \ac{skt}, \ac{emgy}, and \ac{eogy}.
We extracted the same set of features, applied the same classifier (Gaussian \ac{nb}), used the same feature selection, and the same evaluation methodology.

Table~\ref{tab:results-deap} presents the results of our replication, averaged over subjects.
The first row of the table shows the results reported in the original DEAP paper~\cite{deap}, while the second row shows our replication results.
Our replication results are consistent with the original DEAP paper, with no significant differences in performance metrics.
The slight discrepancies observed in the results can be attributed to the details of the signal processing and feature extraction steps, which are difficult to replicate exactly without access to the original codebase and parameters.
We also replicated the baseline voting strategies, with the corresponding values shown in the last three rows of the table. 
These baseline values are identical to those reported in the original work (see Table~7 of Ref.~\cite{deap}).

\begin{table}[!t]
\caption{Results of replicating single-trial emotion classification from the original DEAP paper~\cite{deap}.
Numbers are averaged over subjects. (A: Accuracy)
\label{tab:results-deap}}
\centering
\begin{tabular}{@{}l@{\hspace{2mm}}l|c|l|c|l@{}}
\hline
& & \multicolumn{2}{c|}{\textbf{Valence}} & \multicolumn{2}{c}{\textbf{Arousal}} \\
\cline{3-6}
& \textbf{Setup} & \textbf{A} & \textbf{F1} & \textbf{A} & \textbf{F1} \\
\hline
& Ref. Val.~\cite{deap} & 0.627 & 0.608** & 0.570 & 0.533* \\
& Replication & 0.631 & 0.606** & 0.616 & 0.542* \\
\hline
\multirow{3}{*}{\rotatebox{90}{\scriptsize Baseline\hspace{1pt}}}
& Random & 0.500 & 0.494 & 0.500 & 0.483 \\
& Majority & 0.586 & 0.368 & 0.644 & 0.389 \\
& Ratio & 0.525 & 0.500 & 0.562 & 0.500 \\
\hline
\end{tabular}
\begin{minipage}{\linewidth}
\vspace{0.2cm}
\footnotesize{$F1$-scores are macro-averaged over the high and low classes of the emotional dimensions. Significance levels are determined using a one-sample \textit{t}-test, comparing the $F1$-score distribution to the best value among the baseline voting strategies (*: $p < 0.05$, **: $p < 0.01$.)}
\end{minipage}

\vspace{3em}

\caption{Results of single-trial emotion classification on the EmoWear dataset.
Numbers are averaged over subjects.
(Clf.: Classifier, C-sig.: Cardiac signal (ECG, BVP, SCG), Peri.: Peripheral signals (RSP, ADR, EDA, SKT), A: Accuracy)
\label{tab:results-emowear}}
\centering
\begin{tabular}{@{}r|l|l|l||c|l|c|l@{}}
\hline
& \multicolumn{3}{c||}{\textbf{Setup}} & \multicolumn{2}{c|}{\textbf{Valence}} & \multicolumn{2}{c}{\textbf{Arousal}} \\
\cline{2-8}
& Clf. & C-sig. & Peri. & \textbf{A} & \textbf{F1} & \textbf{A} & \textbf{F1} \\
\hline
1 & NB & ECG & all & 0.576 & 0.554** & 0.603 & 0.542* \\
2 & NB & BVP & all & 0.581 & 0.558** & 0.609 & 0.547* \\
3 & NB & SCG & all & 0.573 & 0.552** & 0.603 & 0.542* \\
4 & SVM & ECG & all & 0.600 & 0.584*** & 0.609 & 0.579*** \\
5 & SVM & BVP & all & 0.595 & 0.585*** & 0.606 & 0.573** \\
6 & SVM & SCG & all & 0.596 & 0.587*** & 0.609 & 0.579*** \\
7 & LR & ECG & all & 0.597 & 0.589*** & 0.608 & 0.577*** \\
8 & LR & BVP & all & 0.600 & 0.591*** & 0.609 & 0.575*** \\
9 & LR & SCG & all & 0.597 & 0.588*** & 0.608 & 0.577*** \\
\hdashline
10 & NB & ECG & RSP & 0.576 & 0.554** & 0.617 & 0.560** \\
11 & NB & BVP & RSP & 0.581 & 0.558** & 0.611 & 0.555* \\
12 & NB & SCG & RSP & 0.581 & 0.558** & 0.625 & 0.570*** \\
13 & SVM & ECG & RSP & 0.600 & 0.584*** & 0.605 & 0.574** \\
14 & SVM & BVP & RSP & 0.595 & 0.585*** & 0.605 & 0.567** \\
15 & SVM & SCG & RSP & 0.595 & 0.585*** & 0.612 & 0.583*** \\
16 & LR & ECG & RSP & 0.597 & 0.589*** & 0.605 & 0.571** \\
17 & LR & BVP & RSP & 0.600 & 0.591*** & 0.605 & 0.569** \\
18 & LR & SCG & RSP & 0.600 & 0.591*** & 0.611 & 0.579*** \\
\hdashline
19 & NB & SCG & ADR & 0.543 & 0.522 & 0.587 & 0.520 \\
20 & SVM & SCG & ADR & 0.572 & 0.561*** & 0.578 & 0.550*** \\
21 & LR & SCG & ADR & 0.575 & 0.564*** & 0.586 & 0.555*** \\
\hline
22 & \multirow{3}{*}{\rotatebox{90}{\scriptsize Baseline\hspace{1pt}}} &
\multicolumn{2}{l||}{Random} & 0.500 & 0.494 & 0.500 & 0.483 \\
23 & & \multicolumn{2}{l||}{Majority} & 0.580 & 0.366 & 0.636 & 0.386 \\
24 & & \multicolumn{2}{l||}{Ratio} & 0.521 & 0.500 & 0.560 & 0.500 \\
\hline
\end{tabular}
\begin{minipage}{\linewidth}
\vspace{0.2cm}
\footnotesize{$F1$-scores are macro-averaged over the high and low classes of the emotional dimensions. Significance levels are determined using a one-sample \textit{t}-test, comparing the $F1$-score distribution to the best value among the baseline voting strategies (*: $p < 0.05$, **: $p < 0.01$, ***: $p < 0.001$.)}
\end{minipage}
\end{table}

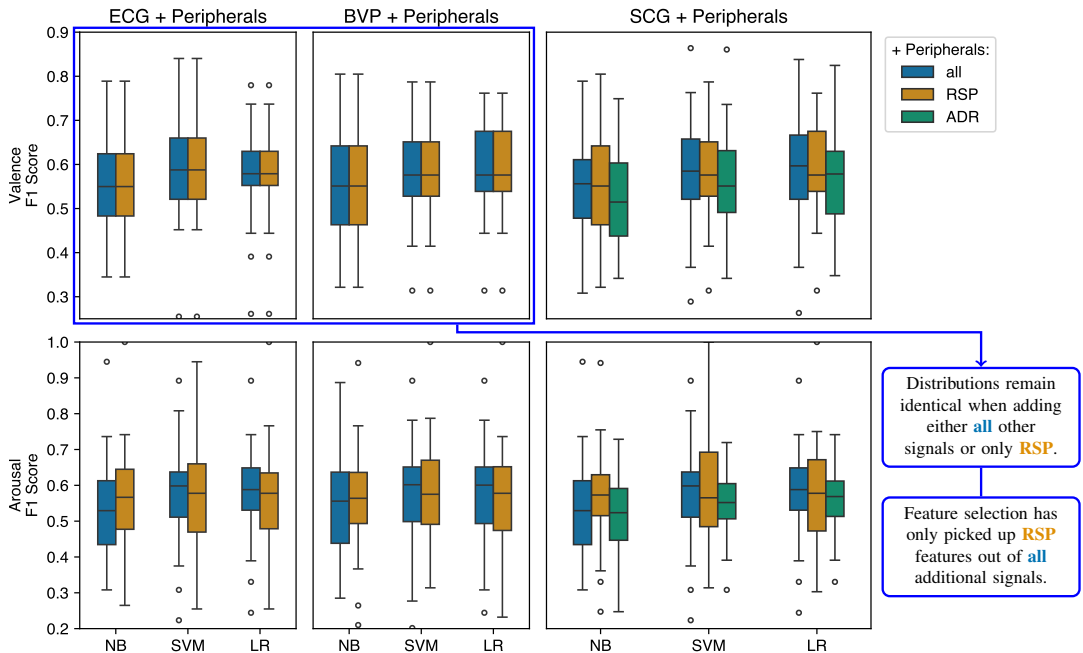
\begin{figure*}[!t]
\centering
\definecolor{colorblind1}{HTML}{0173b2}
\definecolor{colorblind2}{HTML}{de8f05}
\definecolor{colorblind3}{HTML}{029e73}
\adjustbox{width=.8\linewidth}{%
\begin{tikzpicture}
  \begin{pgfonlayer}{background} 
    \node (image) [anchor=south west,inner sep=0] at (0,0) {\includegraphics[width=20cm]{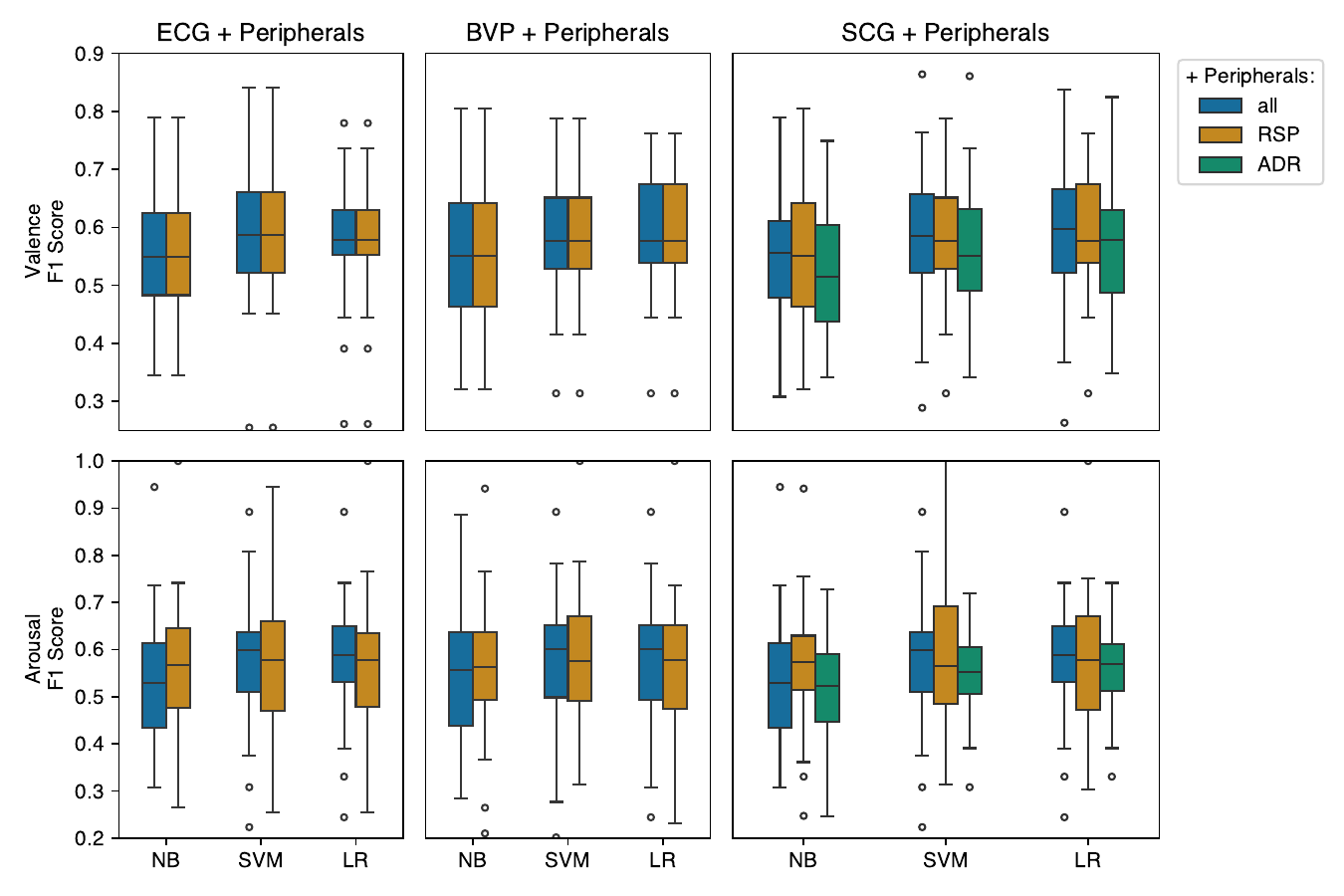}};
  \end{pgfonlayer}
  
  \begin{scope}[x={(image.south east)},y={(image.north west)}]
    \node[anchor=south west] (ecg-bvp) [draw=blue, line width=1.3pt,rectangle,minimum width=9.0cm,minimum height=5.79cm] at (0.082,0.511) {};
    \node (t-01) [rectangle, text=black, text width=3.5cm, align=center, draw=blue,line width=1.3pt,rounded corners,inner sep=5pt,] at (0.97,0.375) {\normalsize Distributions remain identical when adding either \textbf{\textcolor{colorblind1}{all}} other signals or only \textbf{\textcolor{colorblind2}{RSP}}.}; 
    \node (t-02) [rectangle, text=black, text width=3.5cm, align=center, draw=blue,line width=1.3pt,rounded corners,inner sep=5pt,] at (0.97,0.185) {\normalsize Feature selection has only picked up \textbf{\textcolor{colorblind2}{RSP}} features out of \textbf{\textcolor{colorblind1}{all}} additional signals.};
  \end{scope}
  
  \draw [draw=blue, line width=1.3pt,->] ([xshift=3cm]ecg-bvp.south) -- ++(0,-0.15cm) coordinate (mid-point) -| (t-01);
  \draw [draw=blue, line width=1.3pt] (t-01) -- (t-02);
\end{tikzpicture}}
\caption{Boxplots of the $F1$ scores over subjects per classifier type, categorized by the emotional dimensions, and the cardiac sources. Each boxplot corresponds to a cardiac signal that is combined with 1) all other peripheral signals, 2) only the RSP signal, or 3) only the ADR signal.
\label{fig:boxplots}}
\end{figure*}

\subsection{EmoWear Dataset Results}

The EmoWear dataset shares methodological similarities with the DEAP dataset (see Section~\ref{sec:dataset}), enabling us to apply the same emotion classification pipeline.
We kept the processing steps as consistent as possible between the two datasets, making only the following modifications:
1) we extracted additional features from the cardio-respiratory signals (see Table~\ref{tab:features});
2) we added a minimum feature count threshold to the feature selection step (see section~\ref{sec:ml});
3) we tested additional classifiers (\ac{svm} and \ac{lr}); and
4) we examined different combinations of input sets.
All above changes were made to enable a more comprehensive analysis on the EmoWear dataset, especially considering the new gateway modality,~\ac{scg}.

Table~\ref{tab:results-emowear} presents the results of single-trial emotion class\-ification on the EmoWear dataset, averaged over subjects.
The investigated setups can be described as different selections of cardiac signals (\ac{ecgy}, \ac{bvp}, or \ac{scg}), combined with different selections of peripheral signals (either all available signals, or only one of the respiratory signals, \ac{rsp} or \ac{adr}), classified using different models (\ac{nb}, \ac{svm}, or \ac{lr}).
We analyze and discuss these results in the following section.

\section{Discussion\label{sec:discussion}}

\subsection{Methodological Approach}

We based the methodology used in this study, first and most on avoiding common pitfalls, and second on the best practices learned from the literature.
We considered both the limitations and strengths of the EmoWear dataset, such as its limited size, and the availability of the new modality, \ac{scg}.
Based on the lessons learned from the reviewed studies, we ensured:
1) The use of a cross-validation scheme in our data splitting, which is crucial for small datasets to avoid overfitting and to address the generalization of the models over the dataset,
2) The use of $F1$ score as the primary evaluation metrics, which is more robust to class imbalance than accuracy,
3) Averaging the $F1$ scores over both classes, which otherwise risks overestimating the performances by focusing on the $F1$ score of the best performing class,
4) The use of a one-sample \textit{t}-test to assess the statistical significance of the results against some baseline voting strategies, making sure that the results are not due to random chance,
5) Transparency in reporting the methodology and results, which is crucial for reproducibility of the results, and
6) Validating our implementation of the full processing, learning, and evaluation pipeline by replicating the DEAP study results and ensuring consistency.

The consistency of the above methodological approach with the single-trial emotion classification pipeline of the DEAP dataset, allowed us to replicate their work.
Our replication results revealed no significant differences in performance (see Section~\ref{sec:results-deap}).
The consistency of our replication results with the results reported in DEAP study highlights two key points:
1) It serves as a stand-alone contribution by reaffirming their findings, and 
2) It ensures that our implementation of the pipeline is reliable, setting the way for the EmoWear dataset analysis, which we did next.

When comparing  with existing accelerometer emotion recognition research, we note that this study is the first to analyze accelerometer data as a mechanical surrogate of cardiac activity (\ac{scg}) for \ac{er}.
The use of accelerometer data for \ac{er} was previously limited to other motion characterization methods, such as blind motion measurement~\cite{Quiroz2018,schmidt2019multi,Piskioulis2021} or gate-analysis~\cite{Hashmi2020}.
Using \ac{scg} provides a more direct measurement of the heart's mechanical activity, which is less affected by motion artifacts than other accelerometer-based methods.

\subsection{Interpretation of Findings}

\subsubsection*{Comparison to DEAP}
In Table~\ref{tab:results-emowear}, the second row presents the classification results using the \ac{nb} classifier when \ac{bvp} is used as the cardiac signal, and all other peripheral signals present in the EmoWear dataset are used.
As such, this setup is most comparable to the DEAP dataset, which also used \ac{bvp} as the cardiac signal, along with all other peripheral signals, and the \ac{nb} classifier.
Results in this row show applicability of the DEAP setup on the EmoWear dataset, with comparable performance levels.
The level of significance observed in these results are consistent with the reported level of significance in the DEAP study.
It is important to note that we do not expect the results to be identical, as the datasets differ, most importantly the subjects are totally different.
However, the applicability of the DEAP setup to the EmoWear dataset, achieving similar results is a strong indicator, on how well such a setup can generalize across the two datasets.

\subsubsection*{Applicability of SCG for Emotion Recognition}

\begin{figure*}[!t]
\centering
\begin{tikzpicture}
    \node[anchor=south west,inner sep=0] (a) at (0,6) {\subfloat[]{\includegraphics[width=4.35in]{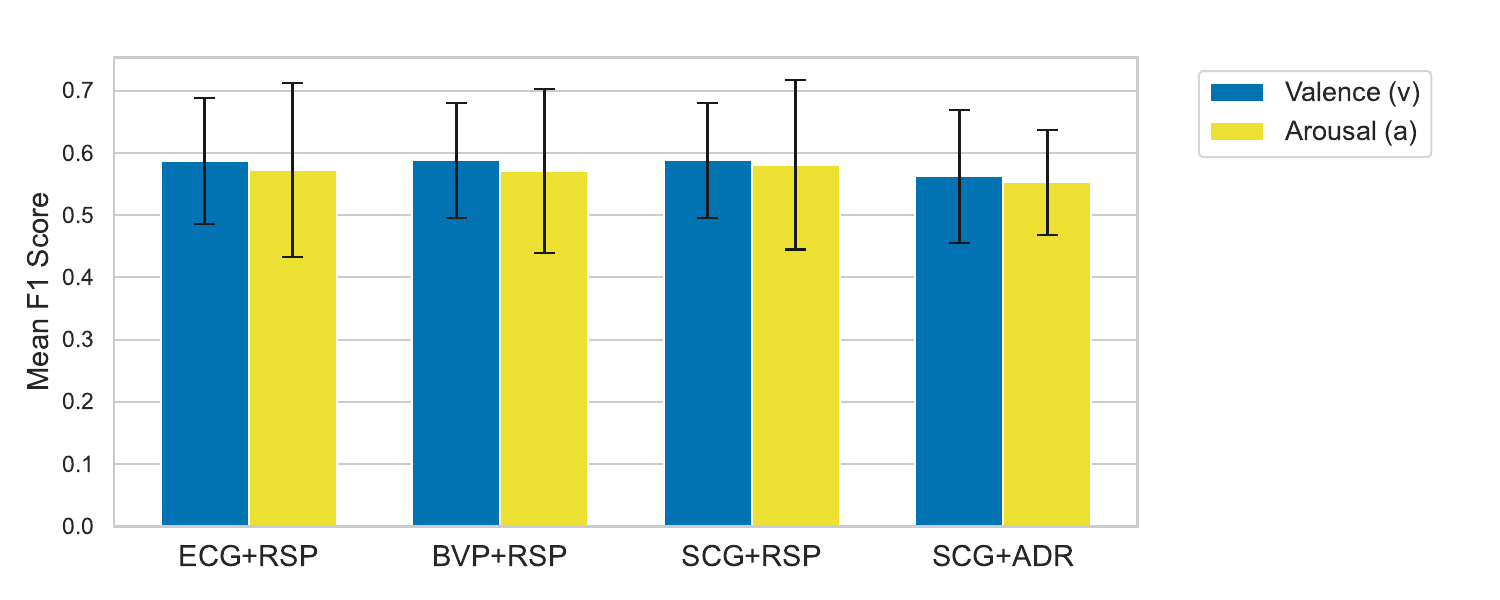}%
    \label{fig:cardio-resp-bars}}};
    \node[anchor=south west,inner sep=0] (b) at (0,0) {\subfloat[]{\includegraphics[width=4.35in]{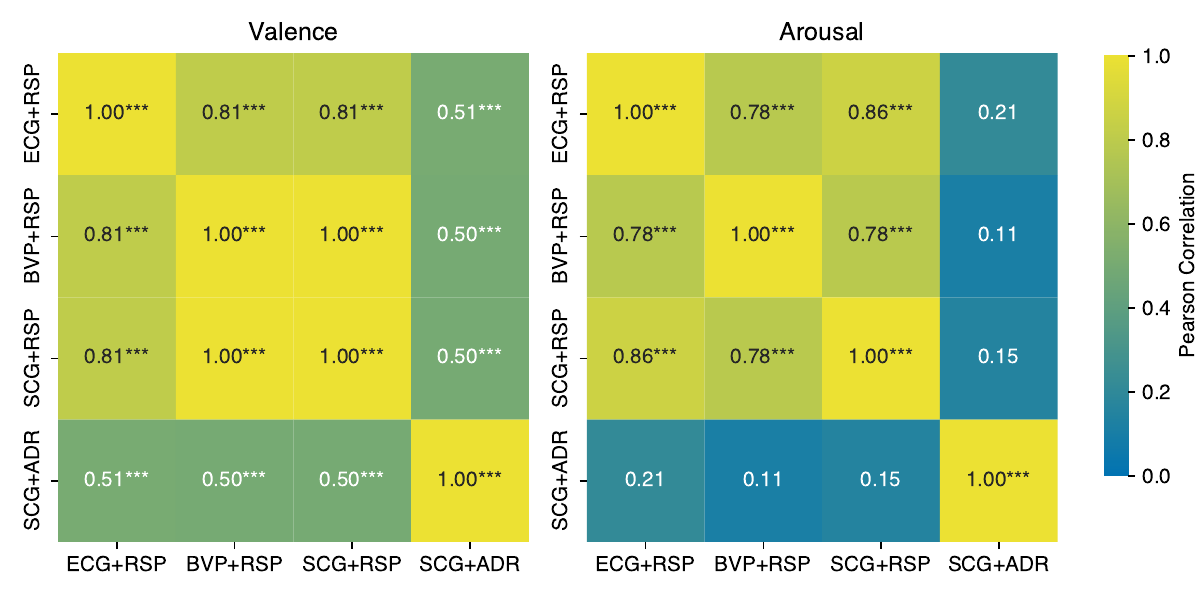}%
    \label{fig:corr-heatmap}}};

    \begin{scope}[x={(b.south east)},y={(a.north west)}]
      \node (a-guide1) [text width=2.3cm,align=center,rectangle,draw=blue,line width=1.2pt,rounded corners,inner sep=5pt] at (1.2,0.77) {\scriptsize SCG+RSP achieves similar mean $F1$ with ECG\,/\,BVP+RSP.};
      \node (a-guide2) [text width=2.3cm,align=center,rectangle,draw=blue,line width=1.2pt,rounded corners,inner sep=5pt] at (1.2,0.61) {\scriptsize SCG+ADR is also comparable but slightly lower.};

      \node (b-annot) [draw=blue,line width=1.2pt,anchor=south west,inner sep=0pt, outer sep=0pt,minimum width=9.65cm, minimum height=1.15cm] at (0.01,0.099) {};
      \node (b-guide) [text width=2.3cm,align=center,rectangle,draw=blue,line width=1.2pt,rounded corners,inner sep=5pt] at (1.2,0.36) {\scriptsize Although SCG+ADR performs differently with other signal pairs, within subjects.};
    \end{scope}

    \draw[->,blue,very thick] ([xshift=-2.7cm]a.east) -- (a.east-|a-guide1.west);
    \draw[blue,very thick] (a-guide1) -- (a-guide2);
    \draw[->,blue,very thick] ([yshift=-0.2cm]b-annot.east) -| (b-guide);

\end{tikzpicture}
\caption{Comparing cardio-respiratory combinations on F1-score results. (a) Averaged over subjects and classifiers (SVM and LR only) with error bars reflecting standard deviations. (b) Pearson correlation heatmap of F1 results obtained (*: $p < 0.05$, **: $p < 0.01$, ***: $p < 0.001$.)
\label{fig:cardio-resp-compare}}
\end{figure*}
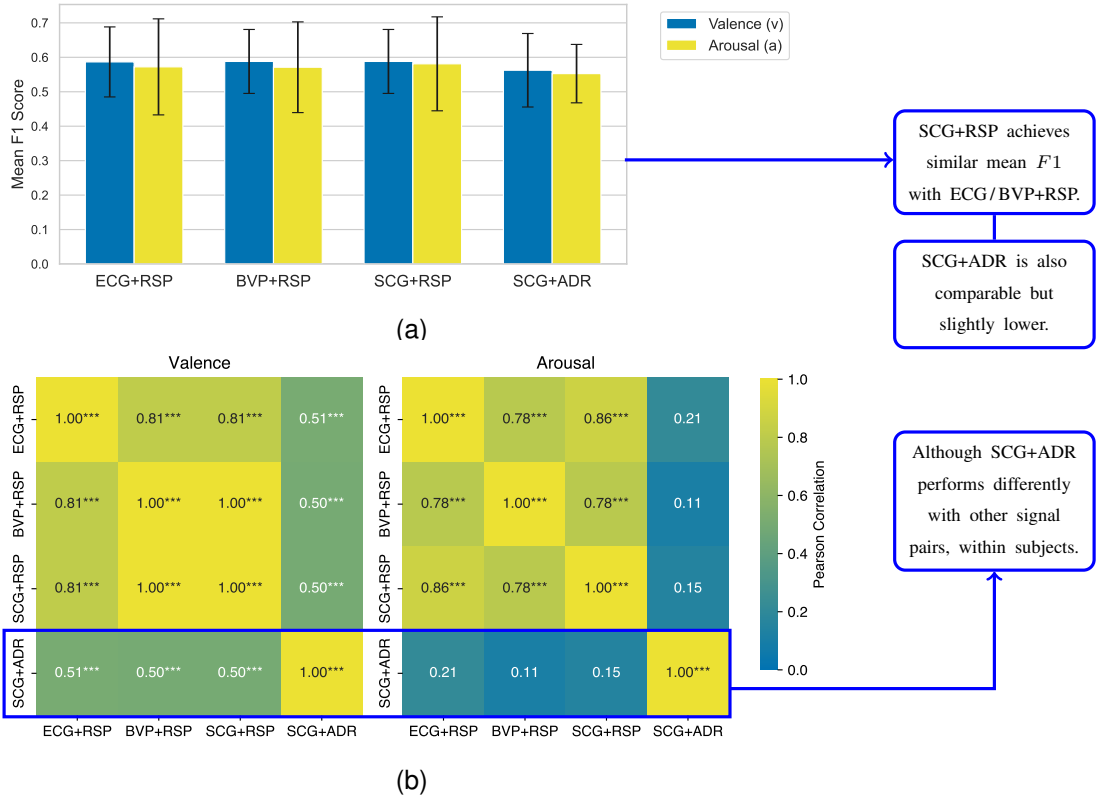

The results of primary interest are those related to the SCG signal, which has never been investigated for \ac{er} before. 
We discuss the outcomes of the \ac{scg}-based setups in three contexts:
1) when combined with all other peripheral signals (\ac{rsp}, \ac{adr}, \ac{eda}, and \ac{skt});
2) when combined with only the respiratory signal from a dedicated sensor (\ac{rsp}); and 
3) when combined with only the respiratory signal from the same sensor that provides the \ac{scg} (\ac{adr}).
Figure~\ref{fig:boxplots} shows distinct boxplots per the three mentioned contexts on the \ac{scg} column.
It also allows for comparing the \ac{scg} results with the other cardiac signals, \ac{ecgy} and \ac{bvp} columns, in the same contexts.

Considering the \ac{scg} in combination with `all' peripheral signals, we observe no systematic trend in the differences of the average results between the \ac{scg} and the other cardiac signals.
The \ac{scg} signal does not outperform the other cardiac signals, but it also does not underperform.
The distribution is obviously different, but so is the distribution of the other two cardiac signals compared to each other.
More or less, similar observations can be made when the \ac{scg} is combined with the \ac{rsp} signal. 

As outlined in the introduction, \ac{er} research relies on mapping objective physiological signals to subjective affective states.
Our findings demonstrate that \ac{scg}, aligns well with this framework, with similar performance to the other cardiac signals.
This is an important finding, as the \ac{scg} signal can be obtained from a simple chest-worn accelerometer, which has the potential to provide a lot more information than just the cardiac signals~\cite{Rahmani2021}.

The most interesting insights are observed when the \ac{scg} signal is combined with the \ac{adr} signal, as this combination uses only a single chest-worn accelerometer to extract both signals. 
The boxplots of the `\ac{scg}\,+\,\ac{adr}' combination in Figure~\ref{fig:boxplots} show an evident drop of performance comparing to their adjacent boxplots, meaning that the \ac{adr} signal has not been able to reach the same level of performance as the \ac{rsp}\,/\,all signals.
Figure~\ref{fig:cardio-resp-compare} provides a more detailed view on the cardio-respiratory-based combination results.
Figure~\ref{fig:cardio-resp-bars} confirms that the average $F1$ scores of the `\ac{scg}\,+\,\ac{adr}' combination is comparable to the `\ac{ecgy}\,/\,\ac{bvp}\,+\,\ac{rsp}' combinations, but slightly lower.
We do not see this performance drop when the \ac{scg} is combined with the \ac{rsp} signal.
Therefore, the slight drop in performance of the `\ac{scg}\,+\,\ac{adr}' combination can be attributed to the fact that the \ac{adr} signal tends to be noisier and less specific than a dedicated respiratory belt (\ac{rsp}), limiting the signal fidelity.
Figure~\ref{fig:corr-heatmap} shows the Pearson correlation heatmap of the $F1$ scores obtained across the four cardio-respiratory signal combinations.
The heatmap suggests that the `\ac{scg}\,+\,\ac{adr}' combination performs differently compared to the other signal combinations, within subjects.
Noteworthily, the different distribution of the `\ac{scg}\,+\,\ac{adr}' combination, does not mean that the combination is not effective.
The \textit{t}-test results in Table~\ref{tab:results-emowear} reveal otherwise:
while the \ac{nb} classifier fails to achieve significant results, the more robust \ac{svm} and \ac{lr} classifiers demonstrate
$F1$ score distributions that are significantly higher than best of the baselines ($p < 0.001$; see Table~\ref{tab:results-emowear}, rows 20--21).
These findings confirm the applicability of a single chest-worn accelerometer for \ac{er}, which is a significant step towards embedding \ac{er} in everyday life scenarios.

\subsubsection*{Classifiers and Features}

The three classifiers used in our study show similar performance trends across the different setups.
However, the \ac{nb} classifier shows lower performance compared to the \ac{svm} and \ac{lr} classifiers.
In Table~\ref{tab:results-emowear}, the \ac{nb} classifier presents the least significant results most frequently.
In the worst case, the \ac{nb} classifier failed to achieve significant results for the `\ac{scg}\,+\,\ac{adr}' combination (row~19), while the \ac{svm} and \ac{lr} classifiers achieved significant results ($p < 0.001$).
Similarly, the boxplots of Figure~\ref{fig:boxplots} suggest consistent results across the \ac{svm} and \ac{lr} classifiers, but slightly lower performance for the \ac{nb} classifier.
Weaker performance of the \ac{nb} classifier was expected, as it comes with strong independence assumptions between features, which do not hold for our extracted features.

Another evident observation in Figure~\ref{fig:boxplots} is the identical distribution of $F1$ scores achieved for valence classification of the `\ac{ecgy}/\ac{bvp}\,+\,all' combinations versus the `\ac{ecgy}/\ac{bvp}\,+\,\ac{rsp}' combinations.
We checked and discovered that our feature selection procedure had only picked up the \ac{rsp} features even when all other peripherals where present.
Such selection was not unexpected, as \ac{skt} contributed minimally to the feature vector~(Table~\ref{tab:features}), and the literature identifies \ac{eda}, the other available alternative, as rather being a primary indicator of emotional arousal~\cite{Delphine2019}.

\subsubsection*{Inter-Subject Variability}

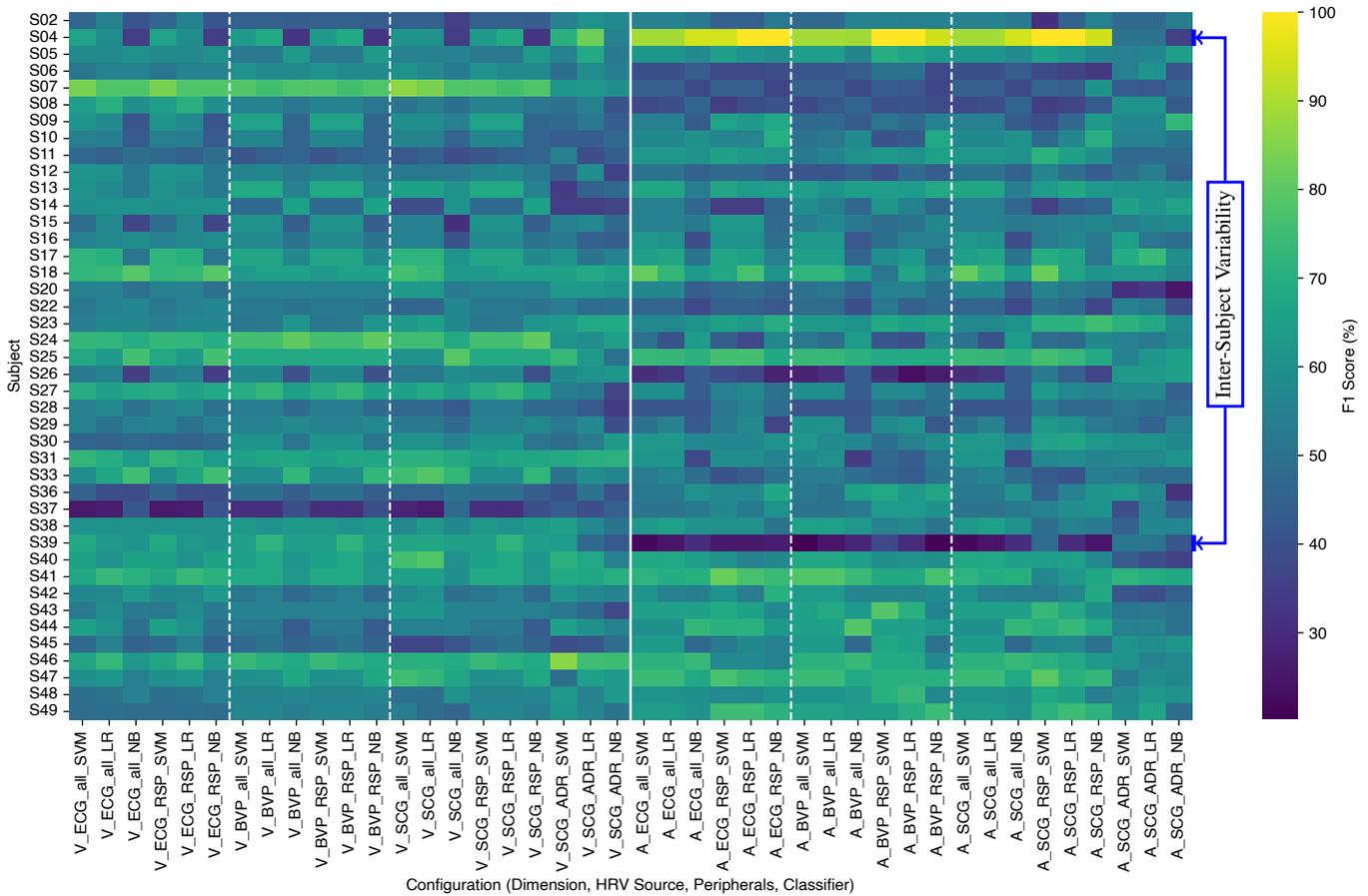
\begin{figure*}[!t]
\centering
\adjustbox{width=0.99\linewidth}{%
\begin{tikzpicture}
  \begin{pgfonlayer}{background} 
    \node [anchor=south west,inner sep=0] at (0,0) {\includegraphics[width=20cm]{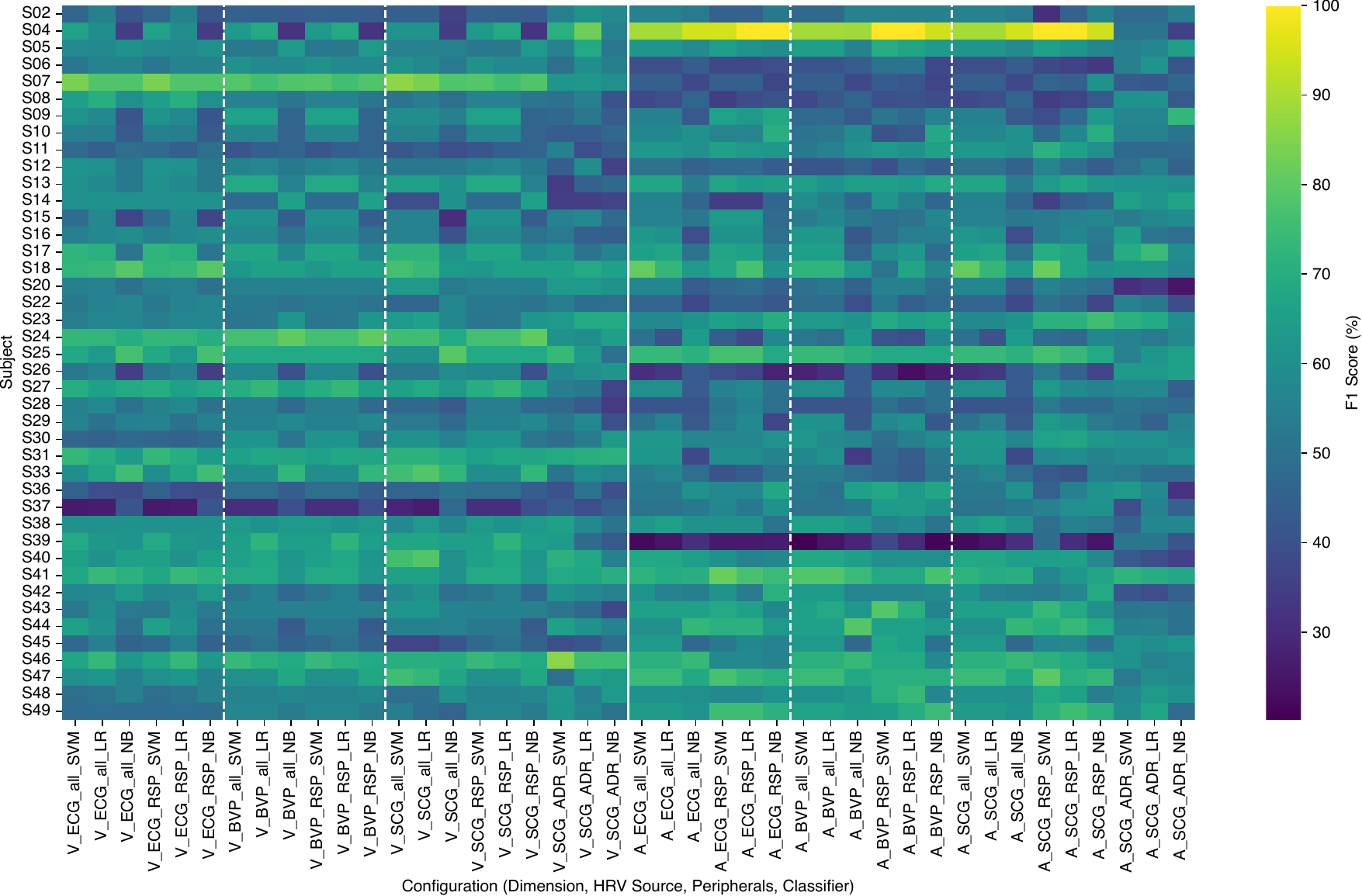}};
  \end{pgfonlayer}

  \node (t-isv) [rectangle, text=black, align=center, rotate=90, draw=blue, line width=1.2pt] at (18.05,8.9) {\small Inter-Subject Variability};
  \node[anchor=south west] (s04a) [draw=blue,line width=1pt,rectangle,minimum height=0.21cm, inner sep=0, outer sep=0] at (17.565,12.60) {};
  \node[anchor=south west] (s39a) [draw=blue,line width=1pt,rectangle,minimum height=0.21cm, inner sep=0, outer sep=0] at (17.565,5.10) {};

  \draw [draw=blue, line width=1.2pt, <-] (s04a) -| (t-isv);
  \draw [draw=blue, line width=1.2pt, <-] (s39a) -| (t-isv);
\end{tikzpicture}}
\caption{Heatmap of all F1 scores per subject, for all configurations of emotional dimension, HRV source, peripherals, and classifiers.
(V: Valence, A: Arousal.)
\label{fig:f1_heatmap}}
\end{figure*}

The variability of the obtained results between subjects is evident in the wide boxplots of Figure~\ref{fig:boxplots}.
Figure~\ref{fig:f1_heatmap} provides a heatmap of all $F1$ scores obtained in the study.
The clear horizontal patterns in the heatmap indicate that performance remains fairly consistent across different setups but varies significantly across subjects.
Some subjects consistently achieve better or worse classification results than others, regardless of the setup 
(although, this pattern is not always observed for the `\ac{scg}\,+\,\ac{adr}' setups.)
The variability is more pronounced in the classification of the valence dimension for the subjects 7, and 37, and in the arousal dimension for the subjects 2, 26, and 39.
This inter-subject variability reaffirms that emotional experience is inherently subjective and can be expressed differently by different individuals~\cite{LeDoux2018}.
We also hope that the detailed heatmap of $F1$ scores in Figure~\ref{fig:f1_heatmap} will serve as a baseline for future studies on the EmoWear dataset, providing further insights into \ac{er} performance across subjects.

\subsection{Limitations}

\subsubsection*{Comparison to Related Work}
In section~\ref{sec:related-work}, we reviewed related work that provide context for our study; however, several limitations prevent direct comparisons between our results and those of the reviewed studies.
First, this study is the first to investigate \ac{er} using the EmoWear dataset, and the first to explore the potential of the \ac{scg} signal for this purpose.
Second, the use of incompatible performance metrics and validation procedures in our studies complicate any direct comparison~\cite{Saganowski2022}.
Finally, methodological gaps that were identified in Section~\ref{sec:related-work} may have led to overly optimistic results in some of the reviewed studies, which we avoided in this study.
However, following the single-trial emotion classification pipeline of the DEAP study~\cite{deap} allowed us to replicate their results and compare our findings.
Moreover, our use of baseline voting strategies provided a reference point for our results, and statistical significance tests confirmed the validity of our findings.

\subsubsection*{Dataset Limitations}
EmoWear and DEAP datasets have inherent limitations that should be considered when interpreting the findings.
First, the demographic composition of the datasets may not fully represent the diversity of real-world users.
Factors such as age, gender, and physiological differences may influence the physiological signals used for \ac{er}.
While the datasets include both male and female participants, a more balanced and heterogeneous sample would be beneficial for ensuring broader applicability.
Second, the experimental setting of the datasets may not entirely reflect real-world usage scenarios.
Although the EmoWear dataset was designed to mimic practical conditions, controlled laboratory settings inherently differ from everyday environments where wearable \ac{er} systems would be deployed.
The impact of movement artifacts, sensor positioning variations, and spontaneous emotional expressions remains an area for further investigation.
Third, both datasets are relatively small, which limits the complexity of the models that can be trained.
Deep learning models require large datasets to avoid overfitting which we address in the following part.

\subsubsection*{Model Complexity and Overfitting}
As previously mentioned in section~\ref{sec:ml}, our selection of the classifiers and their hyperparameters was based on an initial investigation of different classifiers and hyperparameters, aiming to maximize the macro-averaged $F1$ score over the high and low classes of the emotional dimensions in the entire EmoWear dataset.
We experimented with \ac{cnn} and \ac{lstm} models, but they failed to classify emotions effectively, consistent with our expectations based on the dataset size.
Performing Leave-One-Video-Out cross-validation on a subject-dependent \ac{er} task, would leave us with only 37 training samples in each fold which is a very small dataset for deep learning models.
On average over the folds, these models predicted the train data at about 98\% $F1$, while the test data was predicted at a range of about 40-45\%.
Although these models could potentially outperform classic classifiers if trained on a larger dataset, we observed significant overfitting on the training data, even after applying various regularization techniques such as dropout, early stopping, L2 norm regularization, and batch normalization.
We identify a potential direction for future research to address the dataset size limitation, which we discuss later in this section.

\subsubsection*{Subjectivity of Emotions}
Emotions are complex and multifaceted, and can be influenced by a wide range of factors, including cultural background, personal experiences, and individual differences~\cite{LeDoux2018}.
The inherent complexity and subjectivity of emotions is a major challenge in \ac{er}, especially in providing ground truth labels for training and testing the models~\cite{Saganowski2022}.
A manifestation of these issues are reflected in the variability over subjects observed in our results, and discussed earlier.

\subsection{Future Directions}

Outcomes and limitations seen in this study suggest several potential directions for future research.
These directions include but are not limited to:
1) Exploring the potential of the \ac{scg} signal for \ac{er} in more detail, including the development of dedicated feature extraction methods and models;
2) Investigating and improving signal processing techniques for more robust information inference from the \ac{scg} and \ac{adr} signals;
3) Examining the potential of lightweight deep learning models (e.g., MobileNet, TinyLSTM) for \ac{er} on the EmoWear dataset, in addition to the explored traditional classifiers;
4) Exploring deep learning approaches like \acp{cnn}, \acp{lstm}, or transformers to process raw \ac{scg} signals directly, circumventing limitations of manual feature engineering;
5) Exploring transfer learning or self-supervised pre-training on large-scale physiological datasets to overcome the challenge of limited dataset size;
6) Investigating subject-independent \ac{er} problem using stronger models, capable of learning more complex patterns;
7) Uncovering the potential of chest-worn accelerometers for \ac{er} in everyday life scenarios, including the development of real-time \ac{er} systems;
8) Exploring the potential of the EmoWear dataset for other applications, such as voice activity detection, drinking detection, and activity recognition, which are also included in the dataset;
9) Expanding the existing dataset by collecting more data from a larger and more diverse group of participants, to improve the generalizability of the findings; and
10) Developing a chest-worn smartphone application that can be used by participants in their daily lives, to collect more data in an opportunistic manner.

\section{Conclusion\label{sec:conclusion}}

In this study, we explored the potential of \ac{scg} for emotion recognition.
We replicated the single-trial emotion classification pipeline of the DEAP study, applying it on both the DEAP and EmoWear datasets.
Replication results showed consistency with the original DEAP study, both confirming their findings and validating our implementation of the pipeline.
Applying the same validated pipeline to the EmoWear dataset revealed that \ac{scg}
can be used for emotion recognition, achieving similar performance to the \ac{ecgy} and \ac{bvp} cardiac modalities.
Moreover, we achieved significant results using only the combination of \ac{scg} with the \ac{adr}, both extracted from the \ac{acc} signal.
`\Ac{scg}\,+\,\ac{adr}' combination results confirmed that a single chest-worn accelerometer can be effectively used as a physiological gateway for emotion recognition.
Our results serve as the first benchmarks for future research on the EmoWear dataset.
Employing deep learning models was limited by the small size of dataset; however, future work could explore transfer learning or self-supervised pre-training on larger datasets.
Expanding research on \ac{scg} for emotion recognition can bring affective computing technology into everyday life.

\bibliographystyle{IEEEtran}
\bibliography{IEEEabrv,library}

\begin{IEEEbiography}[{\includegraphics[width=1in,height=1.25in,clip,keepaspectratio]{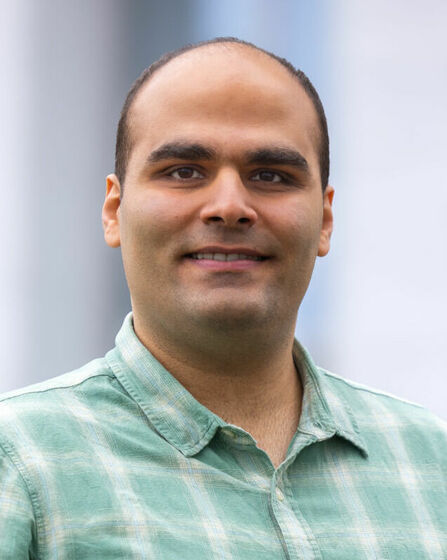}}]{Mohammad Hasan Rahmani}
is currently pursuing a Ph.D. in applied engineering at the University of Antwerp, Belgium, focusing on biomedical signal processing and affective computing.
He received his M.Sc. in Biomedical Engineering from Amirkabir University of Technology (Tehran Polytechnic) and B.Sc. in Electrical Engineering from K.N. Toosi University of Technology, both in Tehran, Iran.
As the principal architect of the ColEmo interface, he developed an open-source tool for collecting emotion data, which was instrumental in creating the EmoWear dataset--a comprehensive resource for emotion recognition and context awareness through wearable sensors. 
He contributed to the NudJIT project, focusing on adaptive interventions to promote healthy behavior through just-in-time nudges.
His research interests include wearable technology, sensing, machine learning, robotics and the integration of physiological signals for enhanced human-computer interaction.
\end{IEEEbiography}

\begin{IEEEbiography}[{\includegraphics[width=1in,height=1.25in,clip,keepaspectratio]{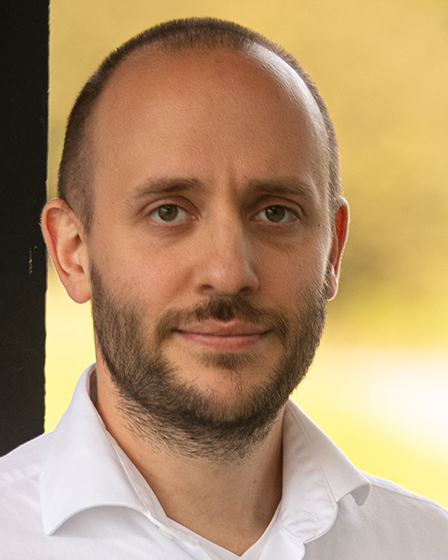}}]{Rafael Berkvens} is an Assistant Professor at IDLab-imec, Department of Electronics, Information and Communication Technology, Faculty of Applied Engineering, University of Antwerp, where he obtained both his Master's and Ph.D. degrees. His research focuses on understanding the information available about individuals and their environment through the study of wireless communication signals. During his Ph.D., he utilized RatSLAM—a computational model of a rodent's brain—replacing camera input with Wi-Fi signals to explore spatial perception. He has developed the concept of opportunistic sensing, aiming to perceive the world using radio frequency transmitters as if they were light sources. He is currently advancing fundamental aspects of this research, including reception, identification, localization, and prediction, with a major focus on Integrated Sensing and Communication (ISAC) and 6G technologies. Prof. Berkvens has published over 90 journal articles and conference proceedings in his field and is a member of the IEEE Communications Society.
\end{IEEEbiography}

\begin{IEEEbiography}[{\includegraphics[width=1in,height=1.25in,clip,keepaspectratio]{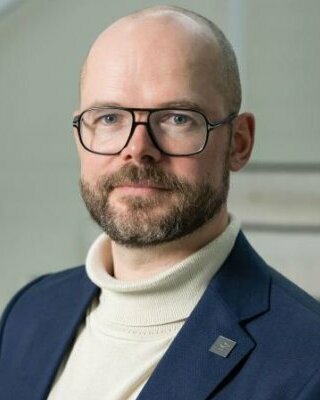}}]{Maarten Weyn} is a full professor and Vice-Rector of Research and Impact at the University of Antwerp. He teaches wireless communication system. His research at imec-IDLab focuses on ultra-low power sensor communication, embedded systems, sub-1 GHz communication, sensor processing, and localization. Maarten co-founded spin-offs Aloxy, CrowdScan, IoSa, and AtSharp, and contributed to 1OK and Viloc.
\end{IEEEbiography}

\vfill

\end{document}